\documentclass[%
 reprint,
superscriptaddress,
nobibnotes,
 amsmath,amssymb,
 aps,
]{revtex4-2}

\usepackage{graphicx}
\usepackage{dcolumn}
\usepackage{xcolor}
\definecolor{myred}{rgb}{1, 0.123, 0.404}
\definecolor{mypink}{rgb}{1, 0.2, 0.745}
\definecolor{mycyan}{rgb}{0.090, 0.667, 0.553}
\definecolor{freshgreen}{rgb}{0.051, 0.796, 0.733}
\definecolor{mygreen}{rgb}{0.184, 0.792, 0.0}
\definecolor{myviolet}{rgb}{0.256, 0.207, 1.00}
\definecolor{myorange}{rgb}{0.9, 0.44, 0.0}
\definecolor{mypink2}{rgb}{0.898, 0.0, 0.451}
\definecolor{mypurple}{rgb}{0.659, 0.251, 1.000}
\usepackage{bm}
\usepackage{hyperref}
\hypersetup{colorlinks, citecolor=mypink,
 linkcolor=myviolet,
 urlcolor=mycyan}
\usepackage{tabularx}
\usepackage{blindtext}
\usepackage{multirow}
\usepackage{amsmath}
\usepackage{adjustbox}

\usepackage{array}

\usepackage[normalem]{ulem} 

\newcommand{\justred}[1]{\textcolor{black}{#1}}
\newcommand{\red}[1]{\textcolor{black}{#1}}
\newcommand{\pink}[1]{\textcolor{black}{#1}}
\newcommand{\blue}[1]{\textcolor{black}{#1}}
\newcommand{\green}[1]{\textcolor{black}{#1}}
\newcommand{\addkl}[1]{\textcolor{black}{#1}}
\newcommand{\fgreen}[1]{\textcolor{black}{#1}}
\newcommand{\purple}[1]{\textcolor{black}{#1}}
\newcommand{\revisecol}[1]{\textcolor{black}{#1}}

\newcolumntype{P}[1]{>{\centering\arraybackslash}p{#1}}

\begin{document}

\preprint{APS/123-QED}

\title{
All-Electron, Density Functional-Based Method for Angle-Resolved \\ Tunneling Ionization in the Adiabatic Regime
}

\author{Imam S. Wahyutama}
\thanks{These authors contributed equally}
\email{wahyutama1@lsu.edu}
\affiliation{Department of Physics and Astronomy, Louisiana State University, Baton Rouge, Louisiana 70803, USA}

\author{Denawakage D. Jayasinghe}
\thanks{These authors contributed equally}
\email{djayas1@lsu.edu}
\affiliation{Department of Chemistry, Louisiana State University, Baton Rouge, Louisiana 70803, USA}

\author{Fran\c{c}ois Mauger}
\affiliation{Department of Physics and Astronomy, Louisiana State University, Baton Rouge, Louisiana 70803, USA}

\author{Kenneth Lopata}
\affiliation{Department of Chemistry, Louisiana State University, Baton Rouge, Louisiana 70803, USA}
\affiliation{Center for Computation and Technology, Louisiana State University, Baton Rouge, Louisiana 70808, USA}

\author{Mette B. Gaarde}
\affiliation{Department of Physics and Astronomy, Louisiana State University, Baton Rouge, Louisiana 70803, USA}

\author{Kenneth J. Schafer}
\affiliation{Department of Physics and Astronomy, Louisiana State University, Baton Rouge, Louisiana 70803, USA}

\date{\today}

\begin{abstract}
We develop and test a method that integrates many-electron weak-field asymptotic theory (ME-WFAT) [\href{https://doi.org/10.1103/PhysRevA.89.013421}{Phys. Rev. A \textbf{89}, 013421 (2014)}] in the integral representation (IR) into the density functional theory (DFT) framework. In particular, we present modifications of the \purple{integral formula in the IR ME-WFAT} to incorporate the potential terms unique to DFT. \blue{\fgreen{By solving an adiabatic rate equation for the angle-resolved ionization yield in our DFT-based ME-WFAT method,}
we show that the results are in excellent agreement with those of real-time time-dependent density functional theory (RT-TDDFT) simulations for NO, OCS, CH$_3$Br, and CH$_3$Cl interacting with one- and two-color laser fields \fgreen{with a fundamental wavelength of $800$ nm}. This agreement is significant because the WFAT calculations take only a small fraction of the time of full TDDFT calculations.}
These results suggest that in the wavelength region commonly used in strong-field experiments ($800$ nm and longer), our DFT-based WFAT treatment can be used to rapidly screen for the ionization properties of a large number of molecules as a function of alignment or orientation between the molecule and the strong field. 
\end{abstract}

\maketitle

\section{Introduction}
Tunnel ionization, wherein an electron driven by an external field moves through the barrier formed by the combination of the molecular and external field potentials, is the ionization process that forms the initial step in many strong field processes. \green{The study of a range of important physical processes in ultrafast science, such as orientation-dependent strong field ionization (SFI), high-harmonic generation \cite{hhg1, hhg2}, the measurement of time delays in tunnel ionization via the attoclock \cite{tunnel_time_delay_he}, light-induced electron diffraction \cite{lied1, lied2}, and ionization-based probes of charge migration \cite{cm_phenyl, cm_adenine, cm_tddft, cm_mode, atto_soliton}, are expected to benefit from an efficient and accurate theoretical method of calculating SFI in a variety of molecules. In light of this overarching influence of tunnel ionization, it is therefore of paramount importance to have theoretical models that can reliably treat this ionization mechanism.}

\blue{Highly accurate SFI yields can be obtained} by propagating time-dependent wave functions or densities, \textit{e.g.} using time-dependent \textit{ab initio} methods \cite{mctdhf_1, mctdhf_2, tdcis, tdcasscf, tdrasscf}, and then recording the amount of density that reaches a pre-determined distance far enough 
\fgreen{from the nuclei to be safely counted as ionized}. 
\blue{However}, the necessity to employ a basis set that spans distances far from the nuclei 
\blue{means that simulation times for such calculations may be prohibitively long}. There have been a number of works focusing on obtaining the ionization probability analytically, \textit{i.e.} without involving wave function evolution, such as the Ammosov-Delone-Krainov (ADK) formulation \cite{adk}, its molecular version \cite{moadk}, the Keldysh-Faisal-Reiss (KFR) model \cite{kfr1,kfr2}, and the weak-field asymptotic theory (WFAT) \cite{wfat1, wfat2, wfat3}, which may be considered as one of the most successful of this kind of approach 
\fgreen{(see Ref. \cite{endo2019angle, wfat_rescattering_co2, wfat_helicity_cf4} for comparison with experiments)}. 

Among a number of different types of WFAT developed over the last decade \cite{wfat1, wfat2, wfat3, tr_mewfat}, \blue{many-electron WFAT (ME-WFAT)} \cite{tr_mewfat} presents several attractive features including that it properly treats the dipole moment during tunneling ionization, uses the actual ionization potential (the difference between neutral and cation eigenvalues) instead of an orbital energy, and is capable of simulating multi-electron effects. 
\fgreen{So far, the ME-WFAT method has been developed in the so-called \textit{tail representation} (TR) \cite{tr_mewfat}, \purple{where the wave functions need to have the accurate asymptotic behavior. This requirement limits the applicability of ME-WFAT to atoms and diatomic molecules because accurate methods of obtaining orbitals having the correct asymptotic are only available for the above types of molecule \cite{hf_grid}.} 
\revisecol{An equivalent formulation in the so-called \textit{integral representation} (IR) \cite{ir_oewfat} removes the above requirement on having the accurate asymptotic, which has also been developed using general Hartree-Fock (HF) framework \cite{ir_oewfat_grid}.}
The reformulation of ME-WFAT within the integral representation makes it capable of further treating tunneling ionization in any molecular geometry. This is presented in Ref. \cite{ir_mewfat} within the \purple{HF framework}, which can be useful for extending IR ME-WFAT to multi-configuration wave functions.
However, given that multi-electron effects are in general treated very efficiently for a wider class of molecules using density functional theory (DFT) than multi-configuration methods, it would be highly desirable to also be able to use DFT wave functions in ME-WFAT calculations.}

In this work, we extend IR ME-WFAT to the framework of 
\blue{DFT} wave functions and demonstrate its ability to reproduce angle-dependent ionization yields obtained by real-time time-dependent density functional theory (RT-TDDFT). The ME-WFAT integral formulae derived using HF orbitals in Ref. \cite{ir_mewfat} are, however, not directly applicable in conjunction with DFT Kohn-Sham orbitals. This is because the 
\fgreen{functional parametrization} in DFT cannot be obtained by starting the analysis from the exact all-electron Hamiltonian without some manual intervention during the derivation. Therefore in this work, we will modify the ME-WFAT integral formulae to accommodate the DFT functionals. We will present the ionization yield calculated using the resulting integral formulae for NO, OCS, CH$_3$Br, and CH$_3$Cl interacting with two-color and one-color lasers 
\blue{with a fundamental wavelength of $800$ nm}, and compare them with RT-TDDFT. 
\fgreen{Here, the time-dependence in our WFAT calculations are emulated by adiabatically solving the exponential rate equation for the yield.}

\purple{We also compare our ME-WFAT results to the one-electron WFAT (OE-WFAT)}
\cite{wfat1, wfat2, wfat3, wfat4, wfat_diatomic_collection, ir_oewfat}. ME-WFAT differs from OE-WFAT in \blue{three main ways}: (i) the dipole moment, (ii) the ionization potential, and (iii) the ionizing orbital. While in ME-WFAT one uses the Dyson orbital, in OE-WFAT, the ionizing orbital must be chosen manually from among the occupied molecular orbitals. 
The ionization potential and dipole moment in OE-WFAT are then taken as orbital energy and dipole moment of the chosen ionizing orbital. The ME-WFAT counterpart of these quantities are taken as the difference between the corresponding neutral and cation values (see Section \ref{sec:theory_mewfat_hf}). This shows that in OE-WFAT, any information about the cation is absent. \fgreen{Not surprisingly,} we will show that ME-WFAT is more accurate than OE-WFAT in many cases presented here. \blue{We note that the three methods compared in this work (ME-WFAT, OE-WFAT, and RT-TDDFT) are only valid for calculations of up to single electron ionization.}

\addkl{We have implemented IR OE-WFAT and ME-WFAT in a development version of the NWChem quantum chemistry package \cite{nwchem}, which allows \purple{for the use of} a wide range of basis sets and DFT exchange-correlation functions.}
\pink{The code is parallelized to handle the computation of some 3D integrals via numerical quadrature, making the simulations scalable to large molecules.}
\addkl{Our WFAT implementation requires few additional specifications beyond a standard DFT input, namely the orbitals of the neutral and cation (for ME-WFAT), the field parameters, and the range of orientation angles.}
\pink{We plan to make our WFAT code publicly accessible by checking into the main branch of NWChem online repository.}

The organization of this paper is the following. In Section \ref{sec:theory_tddft}, we will provide a brief overview of the ionization calculation using RT-TDDFT. In Section \ref{sec:theory_mewfat_hf}, the formulation of ME-WFAT using HF orbitals described in Ref. \cite{ir_mewfat} is summarized. The method proposed here, namely, the ME-WFAT using Kohn-Sham orbitals is outlined in Section \ref{sec:theory_mewfat_dft}, where the modifications needed to make \purple{the HF integral formula} in Section \ref{sec:theory_mewfat_hf} applicable to Kohn-Sham orbitals are presented. Section \ref{sec:theory_oewfat_dft} briefly summarizes OE-WFAT by contrasting it with ME-WFAT in several important aspects. The results are presented in Section \ref{sec:result} where we compare the angle-dependent yields obtained by OE-WFAT, ME-WFAT, and full RT-TDDFT simulations. Finally, future potential improvements of the present method are suggested in Section \ref{sec:conclusion}.

\section{Overview of the Methods} \label{sec:theory}

\subsection{RT-TDDFT with CAP}  \label{sec:theory_tddft}
\blue{Computationally, SFI is a challenging task because one needs to accurately describe many-electron correlation effects while providing a sufficient representation of the continuum electrons \cite{ocs_1}. In this regard, RT-TDDFT provides a balance between the accuracy and speed of the simulation when the system size increases.} In a time-dependent Kohn-Sham (TDKS) framework, RT-TDDFT describes electron dynamics in molecular systems by integrating the TDKS equations,
\vspace{-0.5em}
\begin{align}
\label{eq:TDKS}
  i \frac{\partial \psi (\boldsymbol{r},t)} {\partial t} 
  =&\, 
  \bigg(- \frac{1}{2}\nabla^2 
  + 
  V_{0} (\boldsymbol{r},t) 
  + 
  V_{H}( \rho (\boldsymbol{r},t))    \nonumber\\ 
  &\,
 + V_{\textrm{XC}}( \rho (\boldsymbol{r},t))
  - 
  \boldsymbol{D} \cdot \boldsymbol{E}(t) \bigg) 
  \psi(\boldsymbol{r},t)
\end{align}
where \addkl{$-\frac{1}{2}\nabla^2$} is the kinetic energy \addkl{operator} for the electrons, $V_{0}$ \addkl{contains} the nuclear-electron attraction and the inter-nuclear repulsion, and $\boldsymbol{D} \cdot \boldsymbol{E}(t)$ is interaction of the \addkl{transition dipole matrix} with the applied external field. 
\purple{Unless otherwise mentioned, we use atomic units throughout this work.}
\addkl{In the adiabatic approximation,} the exchange-correlation potential V$_\textrm{XC}$ and the electron mean field repulsion $V_{H}$ depend only on the instantaneous density denoted by $\rho(\boldsymbol{r},t)$ \cite{lopata2011modeling}. \addkl{Typically, RT-TDDFT simulations start from a ground state density matrix converged with DFT, but a superposition of non-stationary density matrices can also be used to emulate a sudden excitation \cite{cm_mode, sideband_hhs, cm_regulation, atto_soliton}.}

\begin{table*}
\caption{A summary of some relevant parameters for the simulations presented in this work. For the functional, hybrid parameters $\alpha_\textrm{RS}$ and $\beta_\textrm{RS}$ ($\beta_\textrm{RS}=1-\alpha_\textrm{RS}$) are dimensionless whereas $\gamma_\textrm{RS}$, the range separation parameter is in units of a.u.$^{-1}$.}

\centering
\vspace{1.0em}
\begin{tabular}{
p{0.1\linewidth} 
p{0.1\linewidth} 
p{0.1\linewidth} 
p{0.28\linewidth} 
r  
P{0.27\linewidth} 
}

Molecule 
& 
{Functional} 
& 
$(\alpha_\textrm{RS},\gamma_\textrm{RS})$ 
& 
Basis Set\footnotemark[1]
& 
{CAP} [{\AA}]\footnotemark[2]
& 
Simulation time\footnotemark[3]
\\
\hline \hline
\vspace{0.2em} & \vspace{0.2em} \\
 
NO & LC-PBE* & $(0.14, 0.55)$ & N,O : aug-cc-pVTZ + ``medium" & 6.5 &    $9.9$h/$0.3$m/$1.1$m @$192$c\\
\vspace{0.01em} & \vspace{0.01em} \\
 
OCS & LC-PBE0* & $(0.0, 0.409)$ & O,C,S : aug-cc-pVTZ + ``medium" & 6.0 &   $28.3$h/$3.4$m/$4.5$m @$144$c\\
 &  &  &  &  & \revisecol{(N.A./$3.4$m/$3.1$m @$144$c)}\\
\vspace{0.01em} & \vspace{0.01em} \\
 
CH$_3$Br & LC-PBE0* & $(0.7,0.3)$ & H,Br : aug-cc-pVTZ & 7.0 &   $23.0$h/$41.7$m/$1.4$h @$144$c\\
 & &  & C : aug-cc-pVTZ + ``large" & & (N.A./$21.1$m/$41.3$m @$288$c)\\
 \vspace{0.01em} & \vspace{0.01em} \\
 
CH$_3$Cl
& LC-PBE0* & $(0.6, 0.15)$ & H,Cl : aug-cc-pVTZ & 7.0  &   (N.A./$6.5$m/$11.5$m @$144$c)\\
& &  & C : aug-cc-pVTZ + ``large" & & \\
 \vspace{0.2em} & \vspace{0.2em} \\
 
\hline

\label{tuned tddft parametes}
\end{tabular}
\footnotetext[1]{``medium" and ``large" refer to the medium and large Schlegel  absorbing basis respectively. The corresponding basis set used by WFAT simulations would be the one where the Schlegel absorbing basis is not included.}
\footnotetext[2]{The distance of the absorbing boundary from the origin.}
\footnotetext[3]{The times are presented in $t_D$/$t_O$/$t_M$ format denoting the simulations times for RT-TDDFT, OE-WFAT, and ME-WFAT, respectively. The suffix "h", "m", and "c" means hours, minutes, and number of cores, respectively. The entries enclosed in parentheses are for simulations with one-color laser (Section \ref{sec:1color}). The ones not inside parentheses are for two-color laser simulations.
\purple{All simulation times are for one Euler angle pair.}
}

\end{table*}

\par
\blue{In this study, we use range-separated hybrid functionals to reduce the self-interaction error and to get the correct long-range Coulomb potential, which is crucial for ionization calculations \cite{tuned_rsx, halomethanes}.}
\addkl{Additionally,} \purple{these functionals} can be tuned by varying both the \addkl{global HF admixture constant} $\alpha_\textrm{RS}$ and \addkl{the range-separation parameter} $\gamma_\textrm{RS}$ \purple{to fulfill} Koopman’s theorem, \textit{i.e.} the energy of the HOMO is equal to the ionization potential of the molecule)\cite{tuned_rsx}.
For all calculations presented here, we tuned both LC-PBE* and LC-PBE0* functionals \cite{lc-pbe0}, which contain pure PBE \addkl{or PBE0} in the short-range and pure HF \addkl{in the long range}. \purple{Such a range-based separation of functionals produces the correct long-range behavior of the effective potential,}\addkl{since the} HF potential is \addkl{asymptotically} the correct $-1/r$.
 
\blue{The way ionization is treated in our implementation of RT-TDDFT comprises two main components, (i) a complex absorbing potential (CAP) placed at some distance from the molecule and (ii) an auxiliary absorbing basis that augments the standard Gaussian one.} 
The CAP distance from the molecule is system-dependent and is chosen such that the ionization rate/yield is insensitive to the CAP position (see Ref. \cite{tuned_rsx} for more details). 
As for the basis, we use the ``medium" and ``large" absorbing bases proposed in Ref. \cite{schlegel_basis} augmented to the standard aug-cc-pVTZ basis. \blue{This auxiliary basis \cite{schlegel_basis} is what allows the density to reach the CAP as a result of ionization. Although these absorbing bases contain highly diffuse Gaussians, the local nature of Gaussian functions demands the CAP to be placed not too far from the molecule. This causes spurious charge removal even when there is no external field. In all TDDFT simulations performed in this work, this error has been corrected for following the procedure outlined in Ref. \cite{tuned_rsx}.}
\addkl{Unfortunately, these large basis sets have many high angular momentum and diffuse functions. Even for moderately-sized molecules, this results in both linear dependencies and significant increases in computational time when evaluating the two-electron integrals. These drawbacks are a strong motivation for finding an  efficient alternative approach to TDDFT when possible.}
The details of tuned functional parameters, CAP positions, and basis sets used in this work are given in Table \ref{tuned tddft parametes}.

\subsection{ME-WFAT using HF orbitals} \label{sec:theory_mewfat_hf}
A full derivation of the angle-dependent ionization yield within the framework of IR ME-WFAT when the wave functions are obtained through the HF method has been given in Ref. \cite{ir_mewfat}. Before presenting the modification required to use these IR ME-WFAT formulas in conjunction with DFT wave functions, we will present a brief overview of WFAT with HF wave functions. \fgreen{Since the HF method has a similar structure to DFT,  knowledge of the form of the IR ME-WFAT integral formula for the HF case can help elucidate the reasons why the corresponding formula for DFT requires additional treatment.}

The derivation of ME-WFAT in either picture starts from the exact $N$-electron Schr\"odinger equation in the presence of a static electric field $\mathbf F=F\hat z$, that is,
\begin{align}
    \label{eq:schroedinger}
     H^{(N)}  \Psi_n(\mathbf X_N) = E_n^{(N)} \Psi_n(\mathbf X_N)
\end{align}
where
\begin{align*}
    H^{(N)} =& \, 
    -\frac{1}{2} \sum_{i=1}^N \nabla_i^2 -
    \sum_{i=1}^N \sum_{I=1}^{N_A} \frac{Z_I}{|\mathbf r_i - \mathbf C_I|}  \nonumber \\
    &\,
  +  \sum_{i=1}^{N-1} \sum_{j=i+1}^N \frac{1}{|\mathbf r_i - \mathbf r_j|} + 
    \sum_{i=1}^N Fz_i,
\end{align*}
$\mathbf C_1, \, \mathbf C_2, \ldots, \, \mathbf C_{N_A}$ are the nuclear coordinates and $Z_1, \, Z_2, \, \ldots, \, Z_{N_A}$ are their charges, and we have used the following notation for the coordinates
\begin{subequations}
   \begin{align*}
       \mathbf X_N &\equiv \{ \mathbf x_1, \ldots, \mathbf x_{N-1},    \mathbf x\}, \\
       \mathbf R_N &\equiv \{ \mathbf r_1, \ldots, \mathbf r_{N-1},    \mathbf r\}, \\
       \mathbf x_i &\equiv (\mathbf r_i, s_i), \\
       \mathbf x &\equiv \mathbf x_N.   
   \end{align*}
\end{subequations}
Here, $\mathbf r_i \equiv (r_i,\theta_i,\varphi_i)$ \purple{in spherical coordinates} and  $s_i$ is the spin coordinate of the $i$-th electron. From this point on, we will assume that the ionized electron has a $\sigma$ spin projection \fgreen{($\sigma=a,b$ with $a=1/2$ and $b=-1/2$)}, and there are $N_\sigma$ occupied orbitals (Kohn-Sham orbitals in the case of DFT) in the neutral. We will also denote 
$\{ \psi_1^a, \ldots, \psi_{N_a}^a, \, \psi_1^b, \ldots, \psi_{N_b}^b \}$ 
and 
$\{ \upsilon_1^a, \ldots, \upsilon_{N'_a}^a, \, \upsilon_1^b, \ldots, \upsilon_{N'_b}^b \}$ to be the occupied spin orbitals in the neutral and cation, respectively.

The underlying idea of ME-WFAT is that in the asymptotic region of one of the electrons, the solution of Eq. \eqref{eq:schroedinger} takes the following \textit{ansatz}:
\begin{align}
    \label{eq:expand_asymptotic}
    \Psi_n(\mathbf{X}_N)\Big|_{\eta\to\infty} = 
    &\,
    \sum_{n'} 
    \Psi_{n'}^+(\mathbf X_{N-1})  
     \nonumber \\
    &\,
  \times  \frac{1}{\sqrt{\eta}}
     \sum_{\nu \sigma}
     f_{\nu\sigma}^{n'n} 
     \mathcal{L}_\nu^{n'n}(\eta,\xi,\varphi) \chi_\sigma(s),
\end{align}
where $|\Psi_{n'}^+\rangle$ is the solution of the Schr\"odinger equation analogous to Eq. \eqref{eq:schroedinger} but defined in the $(N-1)$-electron Hilbert space. $|\chi_\sigma\rangle$ is the spin state having $z$ projection $\sigma$, and lastly $\mathcal L_\nu^{n'n}$ is the one-electron state corresponding to the parabolic channel of $\nu \equiv (n_\xi, m)$ with $n_\xi=0,1,\ldots$ and $m=0, \pm 1, \ldots$. The functional form of $\mathcal L_\nu^{n'n}(\eta=r+z, \xi=r-z, \varphi=\operatorname{arctan}(y/x))$ may be deduced from Ref. \cite{ir_mewfat}.

The expansion coefficient $f_{\nu\sigma}^{n'n}$ in Eq. \eqref{eq:expand_asymptotic} describes the amplitude of the channel connecting a neutral eigenstate $|\Psi_n\rangle$ and another state describing the situation where the $N-1$ electrons around the nuclei occupy a cation eigenstate $|\Psi_{n'}^+\rangle$, whereas the remaining electron is in a spin-parabolic quantum state $|\mathcal L_\nu^{n'n} \, \chi_\sigma \rangle$, and is given by
\begin{align}
    \label{eq:f_coeff}
    f_{\nu\sigma}^{n'n} 
    \approx& \,\, f_{\nu\sigma}^{(0)}(\textrm{IP}, F, \beta, \gamma) \nonumber \\
    =&\,\,
    \sqrt{\frac{\varkappa}{2}}
    \left(
    \frac{4\varkappa^2}{F}
    \right) ^ {\beta_\nu^{(0)}/\varkappa} g_{\nu\sigma}(\beta,\gamma) \nonumber \\
    &\,\,
    \times \exp\left(
    i\frac{\pi}{4} +
    i\frac{\pi\beta_\nu^{(0)}}{\varkappa} -
    \mu_z \varkappa -
    \frac{\varkappa^3}{3F}
    \right),
\end{align}
\green{where 
\fgreen{$(\beta,\gamma)$ are the two Euler angles characterizing the orientation of the molecule relative to the field (a more detailed definition of the orientation angles will be given later in Section \ref{sec:result}). Here, the angle $\beta$ is not to be confused with the spin-down projection defined earlier. Throughout this work, we will make the distinction explicit when the context is ambiguous.} Also, in Eq. \eqref{eq:f_coeff},
$\textrm{IP}=E^{(N)} - E^{(N-1)}$ is the ionization potential, 
$\varkappa=\sqrt{2|\textrm{IP}|}$, $\mu_z$ is the $z$-component in the lab-frame of the vector $\boldsymbol{\mu}^{(N)} - \boldsymbol{\mu}^{(N-1)}$, 
$\boldsymbol{\mu}^{(N)} = -\Big\langle \Psi \Big| \sum_{i=1}^N \hat{\mathbf r}_i \Big| \Psi \Big\rangle$ is the electronic dipole moments of the neutral,
$\boldsymbol{\mu}^{(N-1)} = -\Big\langle \Psi^+ \Big| \sum_{i=1}^{N-1} \hat{\mathbf r}_i \Big| \Psi^+ \Big\rangle$ is the electronic dipole moment of the cation,
$\beta_\nu^{(0)} = Z_c - \varkappa \left( n_\xi + \frac{|m|+1}{2} \right)$ \fgreen{is the adiabatic eigenvalue \cite{wfat1}}, and 
$Z_c = \sum_{I=1}^{N_A} Z_I - N + 1$ is the cation total charge \cite{ir_mewfat}.} 
The approximation in the first line of Eq. \eqref{eq:f_coeff} corresponds to using the leading order approximation (LOA) of WFAT. Because of this approximation, all ground state properties such as wave functions, energies, and dipole moments in the following are associated to the unperturbed systems ($F=0$). 

From this point on, we will omit the eigenstate indices $n$ and $n'$ since HF and DFT produce just one wave function. Here, we follow the authors of Ref. \cite{ir_oewfat_grid} who formulate the \textit{asymptotic coefficient}, $g_{\nu\sigma}$, as
\begin{align}
    \label{eq:asymptotic_coeff1}
    g_{\nu\sigma}(\beta,\gamma) = 
    \sum_{l=|m|}^\infty  \sum_{m'=-l}^l
    I_{\nu lm'\sigma} \, d_{mm'}^l(\beta) \, e^{-im'\gamma},
\end{align}
where $d_{mm'}^l(\beta)$ is the Wigner function and 
\begin{gather}
     I_{\nu lm'\sigma} = \left\langle \Psi^+;\Omega_{lm'}^\nu \chi_{\sigma} \left| \hat V_{1e} + \hat V_{2e}    \right| \Psi \right\rangle, \label{eq:mewfat_integral}  \\
    V_{2e}(\mathbf R_N) = -\sum_{i=1}^{N-1} \frac{1}{|\mathbf r - \mathbf r_i|},  \label{eq:v2e_define}  \\
    V_{1e}(\mathbf r) = \sum_{I=1}^{N_A} \frac{Z_I}{|\mathbf r - \mathbf C_I|}
    -
    \frac{Z_c}{r}.  \label{eq:v1e_define}
\end{gather}
$I_{\nu lm'\sigma}$ is what we will refer to as \textit{ME-WFAT integral}. The function $\Omega_{lm'}^\nu(\mathbf r)$ is the solution of the hydrogen-like Schr\"odinger equation when the energy is not any of the hydrogen-like eigenvalues \cite{ir_oewfat_grid}, and is explicitly given by
\begin{align*}
    \Omega_{lm'}^\nu(\mathbf r) = 
    &\,
    \omega_{\nu l}
    (\varkappa r)^l
    e^{-\varkappa r}
    \nonumber \\
    &\,
    \times M(l+1-Z/\varkappa, 2l+2, 2\varkappa r)\,
    Y_{lm'}(\theta,\varphi).
\end{align*}
where $Y_{lm}(\theta,\varphi)$ is the spherical harmonics and $M(a,b,x)$ is the confluent hypergeometric function \cite{math_handbook_abramowitz} and $\omega_{\nu l}$ is a normalization factor which may be found in Ref. \cite{ir_oewfat_grid}.

 The one- and two-electron parts of the ME-WFAT integral of Eq. \eqref{eq:mewfat_integral} have been derived in Ref. \cite{ir_mewfat} and are given by
\begin{subequations}
    \label{eq:mewfat_integral_term}
    \begin{align}
        \label{eq:v1e_term}
        \Big\langle \Psi^+; \Omega_{lm'}^\nu \chi_{\sigma} 
        \Big|& 
        \hat V_{1e}
        \Big|
        \Psi \Big\rangle = 
        \delta_{M_s'+m_{\sigma},M_s} \left\langle \Omega_{lm'}^\nu \, \Big| \hat V_{1e} \Big| \psi_D^{\sigma} \right\rangle, \\
        \label{eq:v2e_term}
        \Big\langle \Psi^+;\Omega_{lm'}^\nu \chi_{\sigma} \Big|& \hat V_{2e} \Big| \Psi \Big\rangle  \nonumber \\
        =& 
        \,\frac{\delta_{M_s' + m_{\sigma}, M_s}}{\sqrt{N}} \,    (-1)^{N+\delta_{\sigma b}N_a}  \, \nonumber \\
        &\,
        \times \Bigg\{  
        \sum_{k'=1}^{N_{\sigma}-1} \, \sum_{j=2}^{N_{\sigma}} \, \sum_{k=1}^{j-1} 
        \bigg( 
        \left\langle \Omega_{lm'}^\nu \left| \hat J_{k'k}^{\sigma} \right|    \psi_j^{\sigma} \right\rangle \nonumber \\   
        &\,
        -\left\langle \Omega_{lm'}^\nu \left| \hat K_{k'k}^{\sigma} \right|    \psi_j^{\sigma} \right\rangle 
        \bigg) 
        + 
        \big\langle \Omega_{lm'}^\nu \big| \hat{\mathcal{V}}^{\sigma} \big|    \tilde{\psi}^{\sigma} \big\rangle 
        \Bigg\}
    \end{align}
\end{subequations}
\green{
where 
\begin{gather}
    \label{eq:mewfat_j}
    J_{k'k}^{\sigma}(\mathbf r) = \, 
    (-1)^{j+k+k'} 
    \mathcal{R} \mathcal{Q}(k,j,k') 
    V_{k'k}^{\sigma}(\mathbf r), \\
   \label{eq:mewfat_k}
    \Big\langle \mathbf r \Big| \hat K_{k'k}^{\sigma} \Big| \psi_j^{\sigma} \Big\rangle = \,  
    (-1)^{j+k+k'} 
    \mathcal{R}  \mathcal{Q}(k,j,k') 
    V_{k'j}^{\sigma}(\mathbf r) \psi_k^{\sigma} (\mathbf r), \\
    \mathcal{V}^{\sigma}(\mathbf r) =  
    \sum_{k'=1}^{N_{p(\sigma)}}  \sum_{k=1}^{N_{p(\sigma)}} (-1)^{k'+k} \mathcal{S}(k',k) V_{k'k}^{p(\sigma)}(\mathbf r), \\
    V_{k'i}^{\sigma} (\mathbf r) = \, 
    \int d^3\mathbf r' \left(\upsilon_{k'}^{\sigma} (\mathbf r') \right)^*  \frac{1}{|\mathbf r - \mathbf r'|} \, \psi_i^{\sigma}(\mathbf r'), \\
    \label{eq:p_operator}
    p(\sigma) = \,
    \left\{
    \begin{array}{ll}
        a  &, \sigma = b \\
        b   &, \sigma = a
    \end{array}
    \right. .
\end{gather}
$|\psi_D^\sigma \rangle$ is the Dyson orbital corresponding to the removal of an electron from the spin-$\sigma$ channel of the neutral, $\big|\tilde{\psi}^\sigma \big\rangle$ is a vector related to $|\psi_D^\sigma \rangle$, \fgreen{$\mathcal{Q}(k,j,k')$ is the determinant of the overlap matrix in the \textit{ionized spin channel} between the neutral after removing $\{\psi_k^\sigma, \psi_j^\sigma\}$ and the cation after removing $\{\upsilon_{k'}^\sigma\}$, and $\mathcal{R}$ is the determinant of the overlap matrix in the \textit{unionized spin channel} between the neutral and cation.} For more details, we refer the readers to Appendix \ref{app:determinant_coefficients}.} Note that, with Eq. \eqref{eq:p_operator}, one has $N_\sigma + N_{p(\sigma)} = N$.

As has been shown in Ref. \cite{ir_mewfat}, Eq. \eqref{eq:mewfat_integral_term} will reduce to the corresponding equations for OE-WFAT when the cation wave function is formed out of $N-1$ orbitals occupied in the neutral, which is a required property for the formulation so obtained to be mathematically correct. The total angle-dependent ionization rate connecting single determinantal neutral and cation wave functions is then given by
\begin{align}
    \label{eq:iony}
    \Gamma(F,\beta,\gamma) \approx \sum_{\sigma\nu} \, N|f_{\nu\sigma}^{(0)} (\textrm{IP}, F, \beta, \gamma)|^2
\end{align}
accounting for all probabilities for the ionized electron to reside in all possible parabolic states \cite{tr_mewfat}.

\subsection{ME-WFAT using Kohn-Sham orbitals} \label{sec:theory_mewfat_dft}
The reason we need a separate formulation of IR ME-WFAT when the orbitals are of the Kohn-Sham DFT type lies in Eq. \eqref{eq:mewfat_integral}, which only applies to ME-WFAT used in conjunction with Hartree-Fock wave functions. In what follows, we will outline the reason for this and propose some choices that address the freedom that necessarily occurs when using a parameterized method such as DFT.

First, note that the operator part of the bracket in Eq. \eqref{eq:mewfat_integral} lacks the local XC potential term. Second, upon inspection of the two-electron term in Eq. \eqref{eq:v2e_term}, it can be seen that it also does not have the self-interaction (SI) term that makes it possible for the exact exchange in ground state hybrid DFT calculations to depend on the electronic density. Lastly, in hybrid DFT, one needs to multiply the exact exchange with a positive scalar less than unity. This necessitates a suitable identification of the exact exchange part in the right-hand side of Eq. \eqref{eq:v2e_term}. With these three observations in mind, one may see that a brute force application of Eq. \eqref{eq:mewfat_integral} with DFT orbitals will result in a scheme that has little correspondence with TDDFT. And along with the fact that Koopman's theorem for the ionization potential is not satisfied by \purple{approximate} DFT functionals (except for tuned range-separated functionals), the resulting scheme also does not reproduce OE-WFAT under the unrelaxed cation orbital situation mentioned shortly after Eq. \eqref{eq:v2e_term}. 

We would like to note that none of the above modifications are needed in OE-WFAT because the derivation of this method can be initiated from the one-electron Kohn-Sham mean-field equation, hence any DFT-native terms such as the XC potential and the SI term, which are by definition one-particle functions, will automatically be carried over to the OE-WFAT integral formula. This does not apply to ME-WFAT since this method strictly starts off from the exact $N$-electron Hamiltonian.

To begin the analysis, we impose that the ME-WFAT integral formula for use with DFT orbitals should read
\begin{align}
    \label{eq:mewfat_integral_dft}
     I_{\nu lm'\sigma}^{\textrm{M-DFT}} = \left\langle \Psi^+;\Omega_{lm'}^\nu \chi_{\sigma} 
     \left| 
     \hat V_{1e} + \hat V_{2e}^{\textrm{M-DFT}} + \hat V_{\textrm{SI}} + \hat V_{\textrm{XC}}
     \right| 
     \Psi \right\rangle,
\end{align} 
\purple{where $\Psi$ and $\Psi^+$ are the ground state DFT wave functions of the neutral and cation, respectively.} Given the freedom related to the explicit expressions of these missing parts, we choose to use the unrelaxed cation situation mentioned above as a guide for the construction of the following formulas related to the $\hat V_{2e}^{\textrm{M-DFT}}$, $\hat V_{\textrm{SI}}$, and $\hat V_\textrm{XC}$ contributions.

\subsubsection{Local exchange-correlation term}  \label{sec:modify_xc}
\fgreen{The local exchange-correlation potential lies at the heart of practically every DFT calculation, and is a function of electronic density.} Here, within the framework of IR ME-WFAT, we treat this potential as a one-electron potential like the nuclear attraction potential and choose the neutral density (over, for instance, the equally plausible cation density) as the argument of the potential, thus
\begin{align}
    \label{eq:mewfat_xc}
     \left\langle \Psi^+;\Omega_{lm'}^\nu \chi_{\sigma} 
     \left| 
     \hat V_{\textrm{XC}}
     \right| 
     \Psi \right\rangle
     =
     &\,
     \delta_{M_s'+m_{\sigma},M_s} \nonumber \\
     &\,
     \times \left\langle \Omega_{lm'}^\nu \, \Big| \hat V_{\textrm{XC}}(\rho^N) \Big| \psi_D^{\sigma} \right\rangle.
\end{align}
This ensures that this term reduces to the corresponding term in OE-WFAT.

\subsubsection{Self-interaction term} \label{sec:modify_si}
\fgreen{Self-interaction (SI) is a feature of approximate DFT functionals and describes a situation where the electrons are repelling themselves. 
The inclusion of SI, which is largely an unphysical mathematical byproduct of the DFT formalism, in our DFT-based ME-WFAT integral is just to facilitate a comparison with TDDFT in which the SI from the exact exchange term is present. In our implementation of the method in the NWChem package, the inclusion of SI term is optional, and may be omitted if desired.}
To see why SI is missing from Eq. \eqref{eq:mewfat_integral}, we need to look at Eq. \eqref{eq:v2e_term}. Inside the triple sum, had the SI terms existed, we should have terms with $k=j$. Introducing such terms into this sum is, however, not straightforward because $\hat J_{k'k}^\sigma$ and $\hat K_{k'k}^\sigma$ depend on $\mathcal{Q}(k,j,k')$ (see Eq. \eqref{eq:mewfat_j} and \eqref{eq:mewfat_k}), whereas $\mathcal{Q}(k,j,k')$ is undefined when $j=k$ \purple{because in this case, the modified overlap matrix (see Eq. \eqref{eq:define_qcf}) is not square}. We therefore propose to use the following expression for the SI term,
\begin{align}
    \label{eq:mewfat_si_i}
     \Big\langle \Psi^+;\Omega_{lm'}^\nu \chi_{\sigma}
     \Big|&
     \hat V_{\textrm{SI}}^\textrm{type I}
     \Big|
     \Psi \Big\rangle   \nonumber \\
     =&\,
     \delta_{M_s'+m_{\sigma},M_s}
     \int d^3\mathbf r' \left(\Omega_{lm'}^\nu (\mathbf r') \right)^*
     \psi_D^\sigma(\mathbf r') 
     \nonumber \\
     &\,
     \times
     \int d^3\mathbf r \,
     \left|\tilde{\psi}^\sigma (\mathbf r) \right|^2 
     \frac{1-w(|\mathbf r - \mathbf r'|)}
     {|\mathbf r - \mathbf r'|}
\end{align}
where a general weight function
\begin{align}
w(y) = 
    \left\{
    \begin{array}{ll}
        C_X   & \textrm{global exchange} \\
        & \\
        \alpha_{\textrm{RS}} + \beta_{\textrm{RS}} \, \textrm{erf}(\gamma_{\textrm{RS}} \, y)   & \textrm{RS exchange}
    \end{array}
    \right.
\end{align}
has also been inserted to provide applicability of the resulting method to both global and range-separated exchange potentials. The physical motivation behind Eq. \eqref{eq:mewfat_si_i} is that an electron occupying the Dyson orbital representing the ionization channel of interest feels a repulsion field caused by an electron density due to $\tilde{\psi^\sigma}$, which is proportional to the Dyson orbital (see Eq. \eqref{eq:imperfect_dyson} and \eqref{eq:dyson}.

We note that there is not a single way to artificially incorporate SI, for instance, the following expression 
\begin{align}
    \label{eq:mewfat_si_ii}
     \Big\langle \Psi^+;\Omega_{lm'}^\nu \chi_{\sigma} 
     \Big|&
     \hat V_{\textrm{SI}}^\textrm{type II}
     \Big|
     \Psi \Big\rangle   \nonumber \\
     =&\,
     \delta_{M_s'+m_{\sigma},M_s} \frac{(-1)^{N+\delta_{\sigma b}N_a}}{\sqrt{N}} 
     \mathcal{R}
     \nonumber\\
     &\,
     \times \int d^3\mathbf r' \,
     \left(\Omega_{lm'}^\nu (\mathbf r') \right)^*
     \sum_{i=1}^{N_\sigma} 
     (-1)^i \,
     \mathcal{P}(i) \,
     \psi_i^\sigma(\mathbf r') 
     \nonumber \\
     &\,
     \times 
     \int d^3\mathbf r \,
     |\psi_i^\sigma (\mathbf r)|^2 
     \frac{1-w(|\mathbf r - \mathbf r'|)}
     {|\mathbf r - \mathbf r'|},
\end{align}
is equally plausible since, like Eq. \eqref{eq:mewfat_si_i}, Eq. \eqref{eq:mewfat_si_ii} also reduces to the corresponding SI term in OE-WFAT under the unrelaxed cation wave function condition. In all simulation presented in this work, we use the first type of SI term, \textit{i.e.} Eq. \eqref{eq:mewfat_si_i}.

\subsubsection{Exact exchange term}  \label{sec:modify_exact_exchange}
\fgreen{Next, we examine the exact exchange term. }To do this requires an inspection of the algebraic structure of Eq. \eqref{eq:v2e_term}, thus we move the analysis to Appendix \ref{app:exchange_mewfat} and will simply quote the result here. We require that, first, the occupied molecular orbital in the ionized channel corresponding to the largest absolute value of the coefficient $\mathcal{P}$ is moved to the last index, so that after this reordering one has
\begin{align}
    \label{eq:reorder}
    N_\sigma = \operatorname{arg max}_{ i \in [1,N_\sigma]} |\mathcal P(i)|.
\end{align}
Then, the sought expression for the exact exchange is given by
\begin{align}
    \label{eq:v2e_term_dft}
    \Big\langle \Psi^+;\Omega_{lm'}^\nu \chi_{\sigma} \Big|& \hat V_{2e}^{\textrm{M-DFT}} \Big| \Psi \Big\rangle  \nonumber \\
    =& \,\frac{\delta_{M_s' + m_{\sigma}, M_s}}{\sqrt{N}} \,    (-1)^{N+\delta_{\sigma b}N_a}  \, \nonumber \\
    &\,
    \times \Bigg\{  
    \sum_{k'=1}^{N_{\sigma}-1} \, \sum_{j=2}^{N_{\sigma}} \, \sum_{k=1}^{j-1} 
    \bigg( 
    \left\langle \Omega_{lm'}^\nu \left| \hat J_{k'k}^{\sigma} \right|    \psi_j^{\sigma} \right\rangle  \nonumber \\   
    &\,
    - \left\langle \Omega_{lm'}^\nu \left| \hat{\mathcal K}_{k'k}^{\sigma} \right|    \psi_j^{\sigma} \right\rangle 
    \bigg) 
    + 
    \big\langle \Omega_{lm'}^\nu \big| \hat{\mathcal{V}}^{\sigma} \big|    \tilde{\psi}^{\sigma} \big\rangle 
    \Bigg\}
\end{align}
with
\begin{align}
    \label{eq:mewfat_k_dft}
    \Big\langle \mathbf r \Big| \hat{\mathcal K}_{k'k}^{\sigma} \Big|  \psi_j^{\sigma} \Big\rangle 
    =&\, 
    (-1)^{j+k+k'} 
    \mathcal{R}  \mathcal{Q}(k,j,k') \bigg(
    \int d^3\mathbf r'  
    \nonumber \\
    &\, 
    \times 
    \left(\upsilon_{k'}^{\sigma} (\mathbf r') \right)^*
    \frac{w(|\mathbf r - \mathbf r'|)}{|\mathbf r - \mathbf r'|} \, \psi_j^{\sigma}(\mathbf r')
    \bigg) \psi_k^{\sigma} (\mathbf r).
\end{align}
In addition to the attainment of the OE-WFAT formula in the case of the unrelaxed cation, the aforementioned reordering that leads to Eq. \eqref{eq:reorder} \fgreen{also ensures that the total angle-dependent yield  correctly exhibits the symmetry of the molecule when there are multiple channels having the identical ionization potential, \textit{i.e.}, the case of degenerate ionization channels (see Section \ref{sec:ch3br}).}

\subsection{OE-WFAT using Kohn-Sham orbitals}  \label{sec:theory_oewfat_dft}
\blue{
The application of OE-WFAT with DFT Kohn-Sham orbitals is more straightforward than ME-WFAT because the former is constructed starting from a one-electron eigenvalue problem where DFT terms such as the XC functional and the exact exchange are well-defined. In this case, assuming that we choose $\psi_{i'}^\sigma$ as the ionizing orbital, the OE-WFAT integral formula reads
\begin{align}
    \label{eq:oewfat_integral_dft}
     I_{\nu lm'\sigma}^{\textrm{O-DFT}} = \left\langle \Omega_{lm'}^\nu
     \left| 
     \hat V_{1e} + \hat V_{2e}^{\textrm{O-DFT}} + \hat V_{\textrm{XC}}
     \right| 
     \psi_{i'}^\sigma \right\rangle.
\end{align}
where $V_\textrm{XC}(\mathbf r) = V_\textrm{XC}(\rho^N(\mathbf r))$ and 
\begin{align}
    \label{eq:v2e_term_dft_oe}
    \hat V_{2e}^{\textrm{O-DFT}} |\psi_{i'}^\sigma \rangle = 
    \sum_{k=1}^{N_{\sigma}}
       \left(\hat V_{kk}^{\sigma} \left| \psi_{i'}^{\sigma} \right\rangle
       - 
       \hat{\mathcal{K}}_{kk}^{\sigma} \left| \psi_{i'}^{\sigma} \right\rangle \right).
\end{align}
The explicit form of the exchange term (the 2nd term inside the RHS parentheses) in Eq. \eqref{eq:v2e_term_dft_oe} is obtained from Eq. \eqref{eq:mewfat_k_dft} using $j=i'$, $k'=k$ (hence $\upsilon_{k'}^\sigma(\mathbf r) = \psi_k^\sigma(\mathbf r)$), and $\mathcal{R} = \mathcal{Q}(k,i',k) = 1$, see Eqs. \eqref{eq:det_coeff_unrelaxed}. In Eq. \eqref{eq:oewfat_integral_dft}, there is no need to construct a separate SI term because this effect is already contained in $\hat V_{2e}^{\textrm{O-DFT}}$ (the sum over $k$ in Eq. \eqref{eq:v2e_term_dft_oe} includes $k=i'$, which is the source of SI). The other important differences of OE-WFAT from ME-WFAT are the ionization potential and dipole moment, where in the former, the $\textrm{IP}$ is taken as the orbital energy of the $\psi_{i'}^\sigma$ orbital and $\mu_z$ is the lab-frame $z$-component of the dipole moment of this orbital.}
The formulation of OE-WFAT, including the definitions of the IP and the dipole moment, make it clear that OE-WFAT, unlike ME-WFAT, is oblivious to the final state after ionization. 

As a reminder, in arriving at Eq. \eqref{eq:iony} used to calculate the total rate, the channel index (the $(n',n)$ pair) has been dropped due to the use of single determinants in the DFT-based ME-WFAT (see the discussion around Eq. \eqref{eq:f_coeff}). In OE-WFAT where the ionizing orbital is chosen manually (hence, the coefficient $f_{\nu\sigma}^{i'(0)}$ depends on $i'$), it is possible to approximate a given ionization channel starting from the neutral ground state with the ionization from a particular occupied orbital. Therefore, the total rate formula for the DFT-based OE-WFAT is slightly different from Eq. \eqref{eq:iony} only in the presence of an additional sum over $i'$ and in the absence of the prefactor $N$.

\section{Results and Discussions}   \label{sec:result}
We pick four molecules, NO, OCS, CH$_3$Br, and CH$_3$Cl, as test beds to gauge how well the WFAT angle-dependent yield reproduces its TDDFT counterpart. \revisecol{But before presenting the ionization yield of these molecules, in Section \ref{sec:scan_field}, we investigate the intensity dependence of the ionization rates of various channels close to the lowest one in energy (which can be approximated as the ionization from HOMO) to gauge our assumption regarding the importance of the channels other than the lowest one.}

\begin{figure}
    \includegraphics[clip, trim=7cm 13cm 5cm 4cm, width=0.7\linewidth]{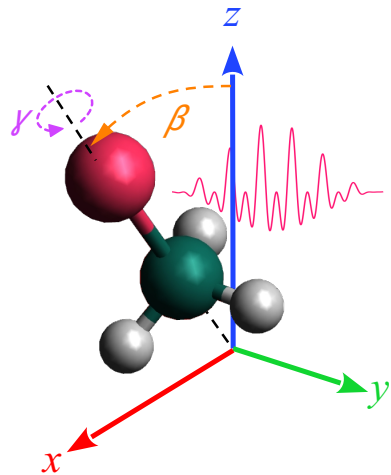}
    \caption{\label{fig:ch3br_angle} An illustration of CH$_3$Br oriented at $(\beta,\gamma)$. $\beta$ and $\gamma$ are the angles of rotation around lab-fixed $y$ axis and body-fixed $z$ axis, respectively. In this example, the body-fixed $z$ axis (not shown) is along the C--Br bond. The arrows associated with each angle denote the rotation direction corresponding to positive value of the respective angles. The two-color laser pulse is also illustrated where its larger oscillations are always in the positive $z$ direction.}
\end{figure}

We use the same laser field for the WFAT and TDDFT simulations for each molecule. In Section \ref{sec:no}, \ref{sec:ocs}, and \ref{sec:ch3br}, this laser is a $(\omega,2\omega)$ two-color pulse with a sine squared envelope, here $\omega=0.057$ a.u. which corresponds to a wavelength of $800$ nm.  The applied external electric field is polarized in $z$ direction and the relative phase is chosen such that its amplitudes are maximized towards positive z direction. \blue{The advantage of using an asymmetric two-color laser is that the directional dependence of the ionization is probed without ambiguity,} \purple{since the ionization signal for a given orientation is mostly due to the field in one direction.}
\justred{In Section \ref{sec:1color}, we use a $\omega$ one-color laser.} The maximum field strength for each molecule is adjusted to be close to the saturation intensity, \pink{and the WFAT and TDDFT yields for a given molecule are normalized so that they have identical maximum yield}. \fgreen{We adjust the laser peak intensity and duration for each molecular system to make the TDDFT computational times tractable and at the same time to have sufficient number of peaks in the field.} Fig. \ref{fig:ch3br_angle} shows how the two orientation angles $0^\circ \leq \beta \leq 180^\circ$ and $0^\circ \leq \gamma \leq 360^\circ$ determine the orientation of the body-fixed axes relative to the lab-fixed axes.

WFAT is, by construction, not a time-dependent method. We can, however, include time dependence quasi-adiabatically by solving the exponential decay rate equation
\begin{align}
    \label{eq:decay_rate}
    \frac{\partial}{\partial t} y(\beta,\gamma,t)
    =
    \Gamma(\mathcal F(t), \beta, \gamma) 
    \left(
    1 - y(\beta,\gamma,t)
    \right).
\end{align}
Here, $\Gamma(\mathcal F(t), \beta, \gamma)$ is the instantaneous total WFAT rate due to a time-varying field $\mathcal F(t)$ obtained from Eq. \eqref{eq:iony} by taking $F = \mathcal{F}(t)$ and $y(\beta,\gamma,t)$ is the quasi-adiabatic yield we are interested in. The actual ionization yield $y(\beta,\gamma,t')$ is then calculated at a time $t'$ when the laser pulse has subsided. \blue{Eq. \eqref{eq:decay_rate} \fgreen{may be expected to be a good approximation} to the actual time-dependent propagation because the field strengths (as confirmed by the amount of ionization) and laser photon energy ($\sim 1.5$ eV) are insufficient to cause significant photoexcitation of the system, hence the ionization can be assumed to proceed adiabatically from the ground state.} 

\begin{figure*}
    \includegraphics[clip, trim=0cm 13cm 2cm 0cm, width=1.0\linewidth]{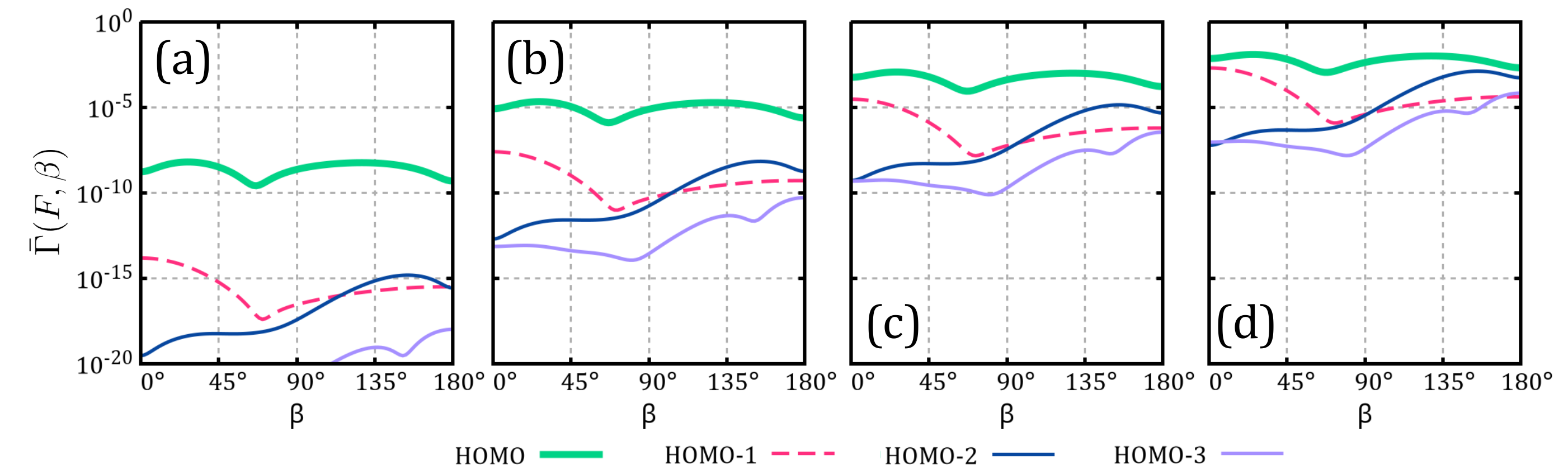}
    \caption{\label{fig:scan_field} 
    \revisecol{
    The average rates, $\bar{\Gamma}_{i'}(F,\beta)$, of some of the high lying occupied molecular orbitals of OCS as a function of field strength. Here $\bar{\Gamma}_{i'}(F,\beta)$ is calculated as the average of 
    $\sum_{\nu\sigma} |f_{\nu\sigma}^{i'(0)} (\epsilon_{i'}, F, \beta, \gamma)|^2$ 
    over $\gamma$, where $i'$ refers to an occupied molecular orbital and $\epsilon_{i'}$ to its orbital energy. The field is static and has a magnitude of (a) $0.02$ a.u., (b) $0.03$ a.u., (c) $0.04$ a.u., and (d) $0.05$ a.u. calculated using OE-WFAT. The vertical axis is in logarithmic scale.}}
\end{figure*}
All WFAT results use the same functional type and parameters as TDDFT as shown in Table \ref{tuned tddft parametes}. The same basis is also used for each molecule except that the Schlegel absorbing basis is removed for WFAT (see Table \ref{tuned tddft parametes}). \purple{Schlegel absorbing basis, which contains very diffuse functions, is not used for WFAT calculations because we find that the presence of too diffuse functions can make the structure factor incorrectly large. This effect is due to the multiplication of the wave functions with $\Omega_{lm'}^\nu$ which increases exponentially with distance \cite{wfat4}.} 
\revisecol{For all simulations here, we use $\nu=\{(0,0), (0,\pm 1), (0,\pm 2), (1,0) \}$ for the summation in Eq. \eqref{eq:iony}.}
The angle-dependent yields presented in the following sections were obtained using the OE-WFAT and ME-WFAT modules implemented in the developer version of NWChem \cite{nwchem}.

\subsection{\revisecol{Field strength dependence of the ionization rates}} \label{sec:scan_field}

\revisecol{
All of the simulations presented in this work assume that the dominant ionization corresponds to the ionization channel connecting the ground states of the neutral and of the cation, which is represented by ionization from HOMO in the OE-WFAT framework. It is therefore instructive to ensure that the intensity of the lasers still falls in the region where higher ionization channels (having higher ionization potentials) are still negligible. For this purpose, we run several static field ionization calculations on OCS with varying field strengths using OE-WFAT.
The resulting $\gamma$-averaged rates are presented in Fig. \ref{fig:scan_field}. The range of field strength used in this result cover the peak field value of the laser used in the subsequent sections. As can be seen from Fig. \ref{fig:scan_field}, up to $F=0.05$ a.u., the rate of ionization from HOMO pretty much still predominates the higher ionization channels (note that the vertical axis is in logarithmic scale). At $F=0.02$ a.u., the rate from HOMO is more than four orders of magnitude larger than HOMO-1. The difference between these channels becomes smaller as the field increases, as expected. In particular, the HOMO rate stays separated from the higher channels while they exhibit some crossings at certain angles even when the field is low. This is because HOMO-1, HOMO-2, and HOMO-3 are relatively close in energy (less than $\sim 1.9$ eV apart) while HOMO is $4.9$ eV above HOMO-1. This observation serves as the justification of our earlier assumption that for the range of intensity used throughout the subsequent sections, only the lowest (HOMO) ionization channel is important.}

\subsection{NO molecule} \label{sec:no}
\blue{We start by calculating the angle-dependent ionization yield in the NO molecule.} The ionization potential as well as the dipole moment components obtained using tuned LC-PBE* and aug-cc-pvtz basis are given in Table \ref{tab:properties}, where the internuclear distance is $1.140$ \AA.
\addkl{Since NO is a radical, an unrestricted Kohn-Sham SCF calculation puts the unpaired electron in a particular 2$\pi$ orbital, which results in an \revisecol{axially} asymmetric charge density. This would cause the ionization yield to vary with $\gamma$ at a given $\beta$. A more physical symmetric density may be obtained by smearing the unparied electron over the two degenerate $\pi$ orbitals. 
Instead of doing that, we use the \revisecol{axially} asymmetric ground state but average over $\gamma$ when computing the yield.
}
\blue{This averaging of WFAT yields is performed formally, that is, for each $\beta$ we sum the yields over $\gamma$ and divide the result by $2\pi$. 
Performing the same averaging for TDDFT yields will, however, be very time-consuming \fgreen{due to the large number of $(\beta,\gamma)$ pairs.} Therefore, for TDDFT yields, the $\gamma$-averaging is performed in the following manner. We first take an arbitrary fixed $\beta$, which is $50^\circ$ for this simulation, and then average the yields over $\gamma$ for this $\beta$. Having obtained this average value, we look for a $\gamma_\textrm{av}$ still with $\beta=50^\circ$ that gives the same ionization yield as the average, and we find that $\gamma_\textrm{av} = 324^\circ$. We then fix $\gamma$ at this value and use it for the scan over $\beta$.} 
\begin{table}[b]
   \caption{\label{tab:properties} Ionization potential (in hartree) and the components of $\boldsymbol{\mu}^{(N)} - \boldsymbol{\mu}^{(N-1)}$ (in a.u.) in body-fixed axes, $\tilde \mu_x$, $\tilde \mu_y$, and $\tilde \mu_z$, used in ME-WFAT for NO, OCS, and CH$_3$Br.}
   \begin{ruledtabular}
      \begin{tabular}{lrrr}
         Properties &
         NO &
         OCS &
         CH$_3$Br \\
         \colrule
         & & & \\ [-0.8em]
         IP               & -0.387291  &  -0.420468  & -0.393985 \\
         $\tilde \mu_x$   &  0.0       &   0.0       &  0.0      \\
         $\tilde \mu_y$   &  0.0       &   0.0       &  0.0      \\
         $\tilde \mu_z$   & -0.016190  &  -0.248725  &  0.811680 \\
         & & & \\ [-0.8em]
      \end{tabular}
   \end{ruledtabular}
\end{table}

\begin{figure}
    \includegraphics[clip, trim=0cm 4.5cm 5.5cm 0cm, width=0.9\linewidth]{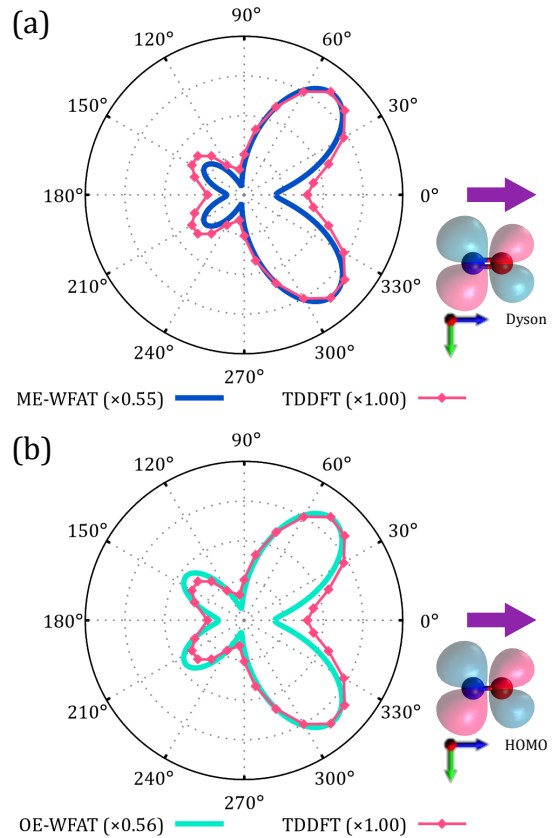}
    \caption{\label{fig:no} NO angle-dependent ion yield \fgreen{resulting from an interaction with a laser pulse having $88$ TW/cm$^2$ maximum intensity ($0.05$ a.u. maximum field) and $22.71$ fs FWHM (for this duration, the saturation intensity calculated following Ref. \cite{halomethanes} is $I_\textrm{sat} = 220$ TW/cm$^2$).} The ME-WFAT and OE-WFAT results are shown in panel (a) and (b), respectively.
    The Dyson orbital and HOMO are also shown in panel (a) and (b), respectively. \blue{The direction of the largest oscillation of the two color field is indicated by the purple arrows near the orbitals, which is fixed in the positive $z$ direction as the molecule rotates (the color of the coordinate axes shown below each orbital has the same meaning as that in Fig. \ref{fig:ch3br_angle}).} The relative orientation between the purple arrow and the orbital indicates its orientation when $\beta=\gamma=0^\circ$. In this schematic, the left atom is N and the right one is O.}
\end{figure}

The ME-WFAT angle-resolved single ionization yield, averaged over $\gamma$, from the NO molecule interacting with a two-color laser having a maximum field of $0.06$ a.u. ($1.26 \times 10^{14}$ W/cm$^2$ in intensity) and \blue{a full width at half maximum (FWHM) duration} of $22.71$ fs
is shown in Fig. \ref{fig:no}(a). It can be seen to exhibit a butterfly-like shape, a characteristic
\fgreen{of the dominant orbital contributing to the ionization having $\pi$ symmetry.}
\fgreen{In ME-WFAT, the Dyson orbital determines the symmetry of the ionization yield, and indeed, in this calculation it has $\pi$ symmetry, see inset in Fig. \ref{fig:no}(a).}
The relatively low ion yields at $\beta = 0^\circ, 105^\circ, 180^\circ, 255^\circ$ can be explained by the nodes of the Dyson orbital. It has a nodal plane containing the molecular backbone and a nodal cone with an apex located between the nuclei. The yield at these nodal angles is, however, nonzero due to the higher parabolic quantum numbers. 

\blue{Fig. \ref{fig:no}(a) also shows the result of TDDFT simulations using identical laser parameters, which are in excellent agreement with the ME-WFAT result}. The butterfly-shaped distributions agree with previous experiments \cite{li2011orientation, endo2019angle}, with the global maxima at $\beta = 50^\circ$ and \revisecol{$\beta=310^\circ$} and  secondary maxima at $142^\circ$ and $218^\circ$, measured with respect to the laser polarization direction. Two asymmetric peaks can clearly be distinguished due to the directional asymmetry of the two-color field. It is also worth noting that these angles correspond to the shape of the 2$\pi$ HOMO as the ionization is preferentially enhanced when the field is directly along a lobe \cite{endo2019angle}.

The angle-dependent yields for ionization from the NO HOMO orbital, computed using OE-WFAT with the same two-color laser, is shown in Fig. \ref{fig:no}(b). 
Here, we see that for NO, ME- and OE-WFAT both yield  good agreement with the TDDFT result. 
\blue{In fact, the shape of the HOMO (inset in Fig. \ref{fig:no}(b)) is very similar to that of the Dyson orbital in panel (a).} 
\revisecol{The most pronounced disagreement between TDDFT and the two WFAT results is the yield at $\beta=0^\circ$. As is suggested by our analysis of the HOMO-1 contribution and the effect of the various parabolic channels, we expect that the inclusion of the first-order correction will bring the shape of the yield around $\beta=0^\circ$ closer to the TDDFT.}
We note that the WFAT and TDDFT results presented in this section for NO agree with experimental data, also obtained using two-color laser fields \cite{endo2019angle}.

\blue{The last column of Table \ref{tuned tddft parametes} compares the simulation times for the TDDFT, OE-WFAT, and ME-WFAT calculations for each molecule. We see that for NO, WFAT simulations are \purple{more than a hundred times} faster than TDDFT.
\purple{We emphasize that the most expensive part of our WFAT algorithm is the calculation of the integrals of Eq. \eqref{eq:mewfat_integral_dft}, but for a given set of molecule and laser parameters these calculations need be performed only once. While the calculation of ionization rate through the use of Eq. \eqref{eq:iony}, \eqref{eq:f_coeff}, and \eqref{eq:asymptotic_coeff1} for all $(\beta,\gamma)$ of interest generally proceeds in just a fraction of the time needed for the preceding integral calculations. So, the times needed to scan all $(\beta,\gamma)$ used in the ionization yield plots is just slightly longer than the WFAT times shown in Table \ref{tuned tddft parametes}. To perform the same $(\beta,\gamma)$ scan in TDDFT would require restarting the TDDFT propagation algorithm for each angle whose total cost is given by the TDDFT timings shown in Table \ref{tuned tddft parametes} times the number of $(\beta,\gamma)$ pairs.}
There are two reasons} \fgreen{why WFAT is much faster than TDDFT.} 
\blue{First, there is no need to have very diffuse functions in the basis set since the method is not based on propagating the wave function or density to a pre-determined, large distance. Second, there is no time evolution algorithm involved. This means that there is no need to repeatedly calculate two-electron integrals needed to construct the Hamiltonian at each time step as in TDDFT.}

\subsection{OCS molecule} \label{sec:ocs}
Carbonyl sulfide (OCS) is a triatomic molecule whose equilibrium geometry is linear. Like NO, its HOMO, which is doubly degenerate, has $\pi$ symmetry. There have been some previous attempts at comparing theoretical angle-dependent yields of OCS with experiments with a varying degree of agreement, some of which used single-color linearly polarized laser fields \cite{ocs_1, ocs_2, ocs_3} while another used a circularly polarized laser \cite{ocs_circular}. 

\begin{figure}
    \includegraphics[clip, trim=0cm 4.5cm 5.5cm 0cm, width=0.9\linewidth]{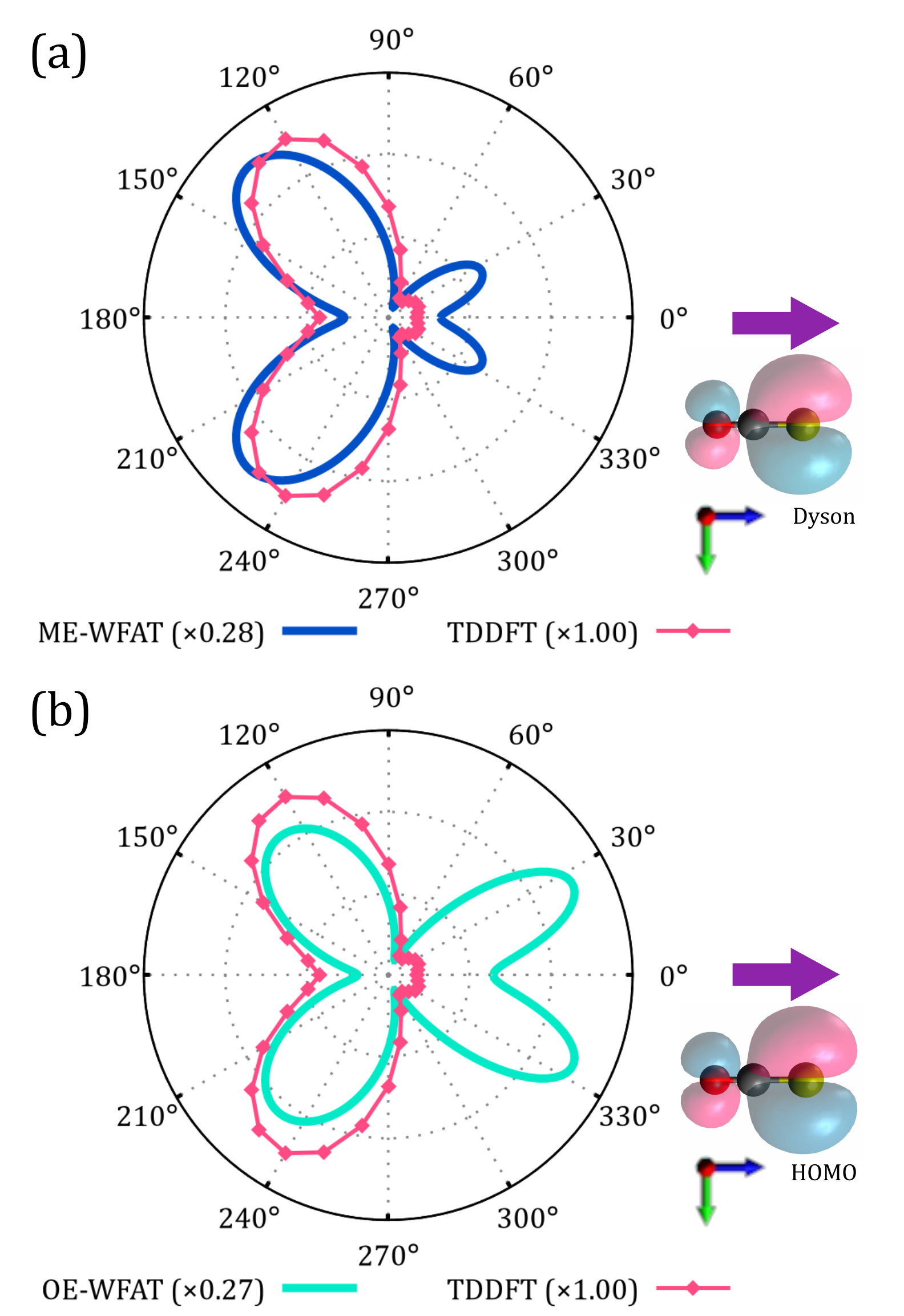}
    \caption{\label{fig:ocs} OCS angle-dependent ion yield comparing (a) ME-WFAT and (b) OE-WFAT with TDDFT results. \fgreen{The laser pulse has $65$ TW/cm$^2$ maximum intensity ($0.043$ a.u. maximum field) and $8.85$ fs FWHM ($I_\textrm{sat} = 176$ TW/cm$^2$).} The meanings of the purple arrow, the orbital image, and the coordinate axes are the same as those in Fig. \ref{fig:no}. In the ball-and-stick representation of the molecule, the left atom is O and the right one is S.}
\end{figure}

The choice of basis and functional mentioned in the beginning of Section \ref{sec:result} produces an ionization potential and dipole moment for OCS given in Table \ref{tab:properties}. The equilibrium geometry for the chosen method and basis occurs at an O-C distance of $1.153$ \AA~and C-S distance of $1.562$ \AA. For this simulation, a maximum field of $0.045$ a.u. ($7.1 \times 10^{13}$ W/cm$^2$) is used with \blue{a FWHM duration} of $8.85$ fs.
\fgreen{The TDDFT yield obtained using these laser parameters is shown with line and markers in Fig. \ref{fig:ocs}, with the maximum found at $\beta = 120^\circ$.} 
At this angle, the laser force component on the electrons along the molecular axis points from O to S.
\addkl{This is consistent with previous experimental and theoretical studies that found a higher ionization rate when the force points from O to S~\cite{ohmura2014molecular}, and that the hole ultimately ends up predominantly on the S atom~\cite{ocs_1}.}

For the ME-WFAT calculation, it is important to note that the single-determinant ground state wave functions of OCS$^+$ are doubly degenerate, which can be associated with the $\pi$-symmetry HOMO of OCS that is also doubly-degenerate. This means one may obtain two different Dyson orbitals (also having $\pi$-symmetry with lobe directions perpendicular to each other) each corresponding to one of the doubly-degenerate cation ground states. By symmetry, the full, unaveraged angle-dependent yields corresponding to these two ``degenerate" Dyson orbitals are trivially connected by a $90^\circ$ shift of $\gamma$. Since the ionization potentials for these two degenerate channels are identical, one should sum the ME-WFAT rates of these two channels to obtain the total rate, which results in a corresponding total yield that is independent of $\gamma$. Before moving on, we would like to emphasize that the neglect of the reordering procedure described in Section \ref{sec:modify_exact_exchange} will result in the absence of this axial symmetry in the full OCS angle-dependent ionization yield.

The ME-WFAT ionization yield (solid line in Fig. \ref{fig:ocs}(a)), is seen to have a reasonably good agreement \fgreen{with the TDDFT yield (line with marker),} with a small difference in the location of the global maxima and in the more pronounced secondary maxima at around $\beta=30^\circ$ and $330^\circ$. 
A possible reason for the latter is the effect of induced core polarization, which has recently been studied in \cite{multielectron_pol_co, multielectron_pol_ocs}. In this mechanism, the external field polarizes the core electron density so that the charge imbalance of the latter creates a counteracting internal field that can substantially decrease the net field for a given orientation, leading to a very low yield around that orientation. \blue{This dynamical effect is not captured by the LOA employed in our ME-WFAT formulation, but can be treated by including the first-order correction to ME-WFAT \cite{tr_mewfat_first_order}.}

The OE-WFAT ionization yield is shown in Fig. \ref{fig:ocs}(b). In contrast to the NO case, the OE-WFAT ionization yield does not agree qualitatively with the TDDFT result due to the peak at $30^\circ$ and $330^\circ$ now becoming the global maximum, although the Dyson orbital and HOMO are still very similar (see the orbital images in Fig. \ref{fig:ocs}). Such a disagreement is caused by the incorrect dipole moment  used by the OE-WFAT to describe the ionization probability connecting the initial neutral state to the final cation state.
\blue{For comparison, the reader is referred to the respective explanations for dipole moment in the paragraphs after Eq. \eqref{eq:f_coeff} and after Eq \eqref{eq:v2e_term_dft_oe}. 
Ref. \cite{ir_mewfat} presents some examples where the main factor dictating the difference of the ME-WFAT and OE-WFAT yields also comes from the shape of the ionizing orbital, \textit{i.e.}, where the Dyson orbital is notably different from the HOMO.}
It is worth mentioning that the solid line in Fig. \ref{fig:ocs}(b) is very similar to the OCS structure factor shown in Fig. 11 of Ref. \cite{wfat4}.
\fgreen{Finally, we note that the computational times for OCS in Table I show that WFAT is, as expected, much faster \purple{(about $150$ times of speed up)} than full TDDFT simulations.}

\subsection{CH$_3$Br molecule} \label{sec:ch3br}

Bromomethane has $C_{3v}$ symmetry and possesses a three-fold rotational symmetry around the C--Br bond. The properties of this molecule relevant to ME-WFAT are given in Table \ref{tab:properties}. The atomic coordinates can be deduced from Table IV of Ref. \cite{halomethanes} since the same basis and functional are used here. 
\begin{figure}
    \includegraphics[clip, trim=0cm 3cm 5.5cm 0cm, width=0.9\linewidth]{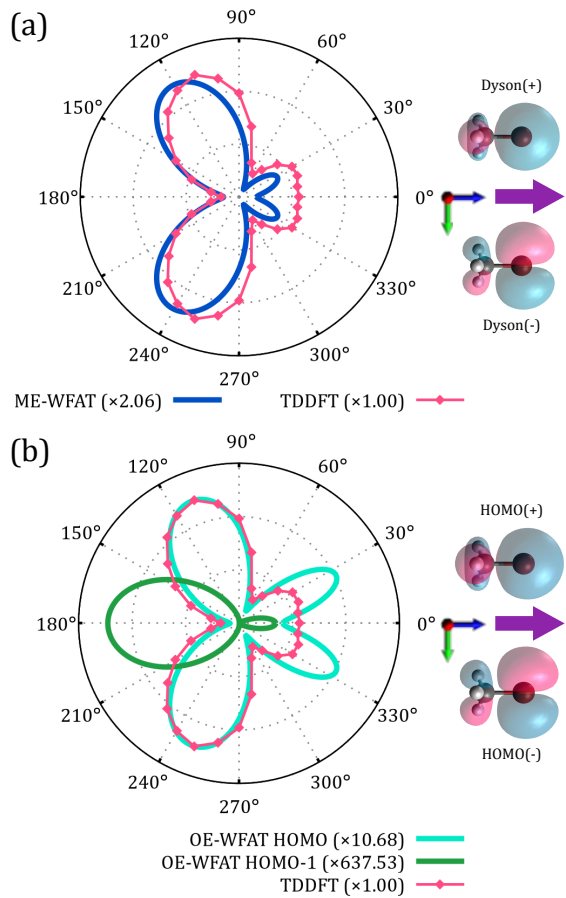}
    \caption{\label{fig:ch3br} CH$_3$Br angle-dependent ion yield comparing (a) ME-WFAT and (b) OE-WFAT with TDDFT results. \fgreen{The laser pulse has $43$ TW/cm$^2$ maximum intensity ($0.035$ a.u. maximum field) and $13.18$ fs FWHM ($I_\textrm{sat} = 150$ TW/cm$^2$).} The meanings of the purple arrow, the orbital image, and the coordinate axes are the same as those in Fig. \ref{fig:no}. In the ball-and-stick representation of the molecule, the right atom is Br and the middle one is C.}
\end{figure}
\blue{
Similar to the case of OCS above, the single-determinant ground state wave functions of CH$_3$Br$^+$ and the HOMO of CH$_3$Br are both doubly-degenerate (see the two orbital pairs in Fig. \ref{fig:ch3br}). This means, the Dyson orbital is also doubly degenerate.
We obtain the doubly-degenerate CH$_3$Br$^+$ ground states by using $N-1$ neutral orbitals, removing one of its two degenerate HOMO's as the starting guess for the cation's SCF iteration.
\revisecol{We identify the two states in these degenerate manifolds (the HOMO and the Dyson orbital) using their reflection symmetry with respect to any of the Br--C--H planes. The one that is even (odd) under this reflection is denoted with the + (--) sign (see the orbital images in Fig. \ref{fig:ch3br}).}
}

\begin{figure}
    \includegraphics[clip, trim=0cm 5cm 6.3cm 0cm, width=0.8\linewidth]{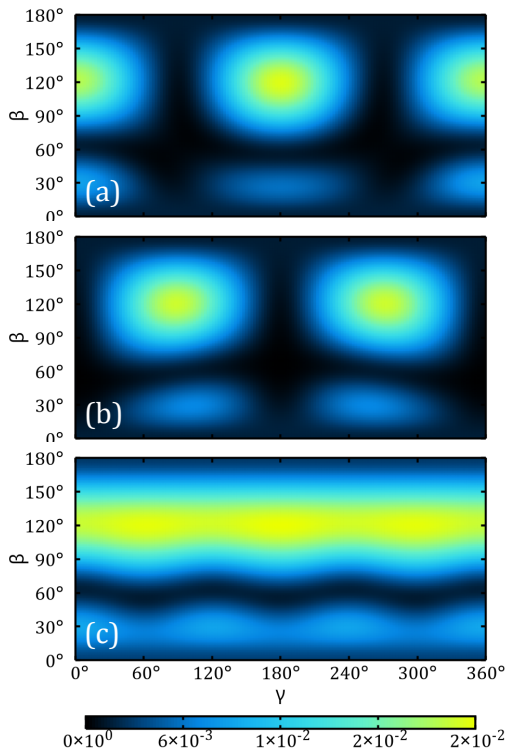}
    \caption{\label{fig:ch3br_2d} The full two-dimensional CH$_3$Br angle-dependent ionization yield. The yield due to each of the doubly degenerate Dyson orbitals are shown in panel (a) and (b). Panel (c) shows the total yield.}
\end{figure}

\blue{
\purple{The full, unaveraged ME-WFAT ionization yield starting from CH$_3$Br in the ground state and ending up in the two degenerate ground states of CH$_3$Br$^+$ are shown in Fig. \ref{fig:ch3br_2d}(a) and (b).
Here, the ionizing field is a two-color laser with a field maximum of $0.05$ a.u. ($8.8 \times 10^{13}$ W/cm$^2$) and an intensity FWHM of $13.18$ fs.
The total yield, shown in Fig. \ref{fig:ch3br_2d}(c), is the sum of the yields corresponding to the degenerate cation final states in Fig. \ref{fig:ch3br_2d}(a) and \ref{fig:ch3br_2d}(b).}
Here, we observe a $120^\circ$ periodicity along the $\gamma$ axis of the yield, which is a reflection of the three-fold rotational symmetry of CH$_3$Br around the C-Br bond. Again, the neglect of the reordering procedure described in Section \ref{sec:modify_exact_exchange} will result in the absence of the three-fold rotational symmetry seen in Fig. \ref{fig:ch3br_2d}(c).}

\blue{
The $\gamma$-average of the yield in Fig. \ref{fig:ch3br_2d}(c) is shown in Fig. \ref{fig:ch3br}(a) as solid line. For the TDDFT yield, the same averaging method as in Section \ref{sec:no} for NO is also employed here, where we set $\beta=120^\circ$, perform an average over $\gamma$, and find $\gamma_\textrm{av}$ that gives identical ionization yield as this average. This procedure yields $\gamma_\textrm{av} = 36^\circ$ as the value to be used for the subsequent scan over $\beta$. A comparison of ME-WFAT (solid line) and the TDDFT yields (dashed line) in Fig. \ref{fig:ch3br}(a) tells us that ME-WFAT is again a reliable alternative to TDDFT for calculating angle-resolved ionization. The ME-WFAT yield in Fig. \ref{fig:ch3br}(a) also looks similar to the result of a previous work on CH$_3$Br interacting with a static field simulated using time-dependent configuration interaction \cite{schlegel_ch3x}.}

\blue{
The angle-dependent yield obtained using OE-WFAT is shown in Fig. \ref{fig:ch3br}(b) as thick solid line. The shape of the maxima peaking at $111^\circ$ and $249^\circ$ have a better agreement with those of TDDFT, but the strength of the secondary maxima at $26^\circ$ and $334^\circ$ is almost twice as large as the corresponding TDDFT values.}

\revisecol{Fig. \ref{fig:ch3br}(a) reveals that the biggest discrepancy between TDDFT and ME-WFAT yields lies around $\beta=0^\circ$. We identify two possible reasons to this: (1) the contribution of the ionization channel connecting the neutral ground state to the cation first excited state (or in OE-WFAT framework, the contribution of HOMO-1), and (2) the orbital distortion effect. The first possibility, however, cannot be investigated using ME-WFAT with DFT orbitals since the excited state wave function requires a multi-determinant description. Instead, we use OE-WFAT using HOMO-1 orbital as the ionizing orbital to shed some light, albeit less accurately.}
\blue{
We identify that HOMO-1 has a $\sigma$-like character.
The resulting ionization yield is shown in Fig. \ref{fig:ch3br}(b) denoted as HOMO-1 in which we see that the yield at $\beta=0^\circ$ is nonzero for this contribution. Note that the relative magnitude between the yields at $\beta=0^\circ$ and at $\beta=180^\circ$ in the HOMO-1 contribution is the opposite to that in the TDDFT yield which is greater at $\beta=0^\circ$. 
\revisecol{
Also, the difference in the magnitude between HOMO and HOMO-1 (compare the scaling factors) looks to be too big to warrant noticeable effect of the latter around $\beta=0^\circ$.}
We attribute these seemingly contradicting behavior to the inaccuracy of OE-WFAT.}

\revisecol{
The second possibility, the orbital distortion due to the laser, is not possible to simulate using the current version of DFT-based ME-WFAT as it requires extending the current formalism to include the first-order correction.}
Aside from this small difference at $\beta=0^\circ$, the overall shape of the ME-WFAT angle-dependent yield agrees qualitatively with TDDFT.

\subsection{One-Color Ionization from \revisecol{OCS}, CH$_3$Cl, and CH$_3$Br} \label{sec:1color}

\begin{figure}
    \includegraphics[clip, trim=0cm 14cm 7cm 1cm, width=0.9\linewidth]{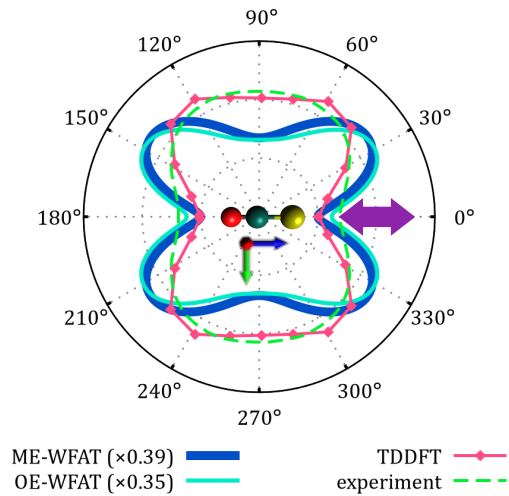}
    \caption{\label{fig:1color.ocs}
    \revisecol{OCS angle-dependent ionization yields due to a one-color laser. The peak (saturation) intensity is $70$ TW/cm$^2$ ($80$ TW/cm$^2$), with a FWHM duration of $37$ fs. The relative orientation between the molecule, the laser field polarization (purple arrow), and the coordinate axes as shown indicates the orientation when $\beta=\gamma=0^\circ$. Additionally, TDDFT calculations and experimental measurement extracted from Ref. \cite{ocs_1} are also plotted. The WFAT values are scaled so that their maxima coincide with the maximum of the TDDFT yield.}
    }
\end{figure}
In this section, we provide further results demonstrating the capability of ME-WFAT to efficiently reproduce the more accurate TDDFT calculations and available experimental data. \revisecol{Fig. \ref{fig:1color.ocs} shows the ionization yield for OCS interacting with a one-color sine-squared pulse having a wavelength of $800$ nm and an intensity FWHM of $37$ fs obtained by ME-WFAT, OE-WFAT, TDDFT, and from experimental data extracted from Ref. \cite{ocs_1}. The peak intensity is $70$ TW/cm$^2$. For this case, we observe a notable discrepancy between WFAT and TDDFT or experiment. We attribute this to the difference in the loci of the global maxima in Fig. \ref{fig:ocs}(a) for ME-WFAT and to the difference of the yield values around $\beta=30^\circ$ and $\beta=330^\circ$ in Fig. \ref{fig:ocs}(a) and (b) for both WFAT results. We expect that that this disagreement in the case of OCS will be remedied by the inclusion of the first-order correction to WFAT.}

\begin{figure}
    \includegraphics[clip, trim=0cm 2cm 3cm 0cm, width=0.95\linewidth]{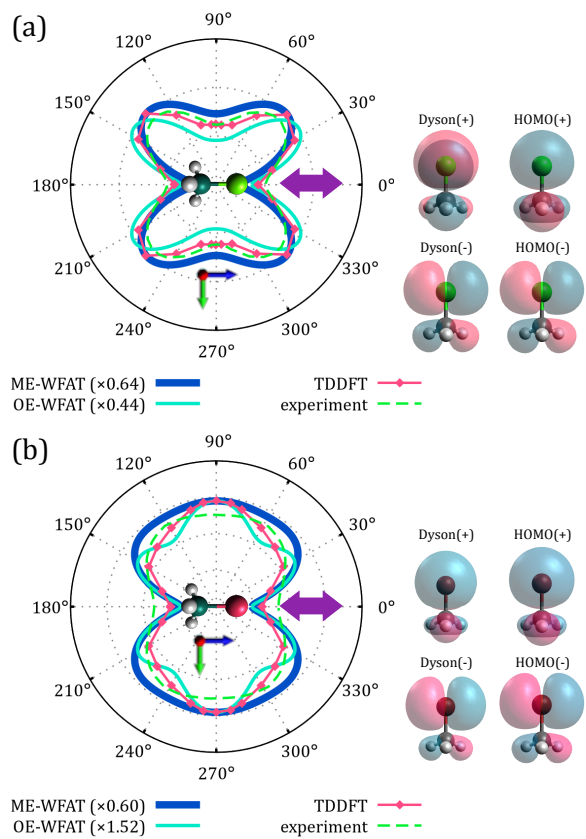}
    \caption{\label{fig:1color.oe_vs_me}(a) CH$_3$Cl and (b) CH$_3$Br angle-dependent ionization yields due to a one-color laser. \fgreen{The peak (saturation) intensity for (a) is $64$ TW/cm$^2$ ($80$ TW/cm$^2$), whereas for (b), it is $60$ TW/cm$^2$ ($50$ TW/cm$^2$), with a FWHM duration of $37$ fs for both.} In the ball-and-stick representation of the molecule, the right atom is the halogen. The relative orientation between the molecule, the laser field polarization (purple arrow), and the coordinate axes as shown indicates the orientation when $\beta=\gamma=0^\circ$. \red{Additionally, TDDFT calculations and experimental measurement extracted from Ref. \cite{halomethanes} are also plotted.} \revisecol{The WFAT values are scaled so that their maxima coincide with the maximum of the TDDFT yield. The Dyson orbitals (used in ME-WFAT) and HOMO (used in OE-WFAT) are also shown.}}
\end{figure}

 Fig. \ref{fig:1color.oe_vs_me} shows the angle-dependent ionization of (a) CH$_3$Cl and (b) CH$_3$Br interacting with a one-color laser \revisecol{having the same parameters as those used to obtain Fig. \ref{fig:1color.ocs} except that the peak intensity for panel (a) is $64$ TW/cm$^2$ and for panel (b) it is $60$ TW/cm$^2$. The experimental data is extracted from Ref. \cite{halomethanes}.} The range-separated functional parameters and basis used in Fig. \ref{fig:1color.oe_vs_me}(a) are given in Table \ref{tuned tddft parametes}.

As shown in Fig. \ref{fig:1color.oe_vs_me}, ME-WFAT yields produce a much better agreement with TDDFT as well as experiment compared to OE-WFAT results. In particular, for CH$_3$Cl, the peaks in both TDDFT and ME-WFAT within the range of $0 \leq \beta \leq 180^\circ$ are located at $45^\circ$ and $135^\circ$, whereas in the same $\beta$ interval, OE-WFAT produces peaks at $35^\circ$ and $145^\circ$. In the case of CH$_3$Br, both TDDFT and ME-WFAT yields have a smooth shape in the range of $0 \leq \beta \leq 180^\circ$ while OE-WFAT features three separated peaks. The peaks at $33^\circ$ and $147^\circ$ are mainly contributed by the secondary maxima around the same angle seen in \fgreen{the OE-WFAT(HOMO) yield in} Fig. \ref{fig:ch3br}(b).

\revisecol{The qualitative difference in the shape of the yields between CH$_3$Cl and CH$_3$Br, especially the fact that at $90^\circ$ and $270^\circ$ CH$_3$Br features maxima while CH$_3$Cl features minima, can be related to the shape of the orbitals. In particular, we see from both the HOMO and Dyson orbitals of the two molecules that the relative electron density in the region around the halogen to that around the hydrogens is higher when the halogen is Br rather than Cl.}

\section{Conclusion and Outlook} \label{sec:conclusion}
In this work, we \fgreen{have demonstrated} that ME-WFAT in the \addkl{leading order approximation (LOA)} is a promising alternative which, when combined with a quasi-adiabatic treatment of the electric field using Eq. \eqref{eq:decay_rate}, produces an efficient and accurate method for calculating tunnel ionization in the adiabatic regime. We emphasize that the application of ME-WFAT for the general molecular case as presented here is made possible by reformulating the tail representation (TR) \cite{tr_mewfat} in the \addkl{integral representation (IR)} \cite{ir_mewfat}. In the IR, one can bypass the need for having the accurate exponentially decaying tail of the orbitals, which is only reliably possible for atoms and diatomic molecules \cite{hf_grid}.

We have applied the LOA of IR ME-WFAT that has been derived in Ref. \cite{ir_mewfat} for Hartree-Fock orbitals to the case of (DFT) Kohn-Sham orbitals. The use of Kohn-Sham orbitals in ME-WFAT requires modification of the ME-WFAT formula to include missing terms such as the XC potential and the self-interaction terms, as well as identifying a suitable exact exchange term. \purple{Several reformulation are possible} in the expression of each of the aforementioned terms, which can be attributed to the parametrization of DFT. By comparing the ME-WFAT against TDDFT angle-dependent yields, we showed that the particular choice of these expressions proposed in this work is indeed suitable.

\fgreen{We have chosen to study the molecules} NO, OCS, CH$_3$Br, and CH$_3$Cl interacting with either one-color or two-color laser pulses to test the reliability of ME-WFAT as an alternative method for the calculation of tunnel ionization. In all these cases, we showed that ME-WFAT is able to reproduce TDDFT angle-dependent ionization yields, whereas OE-WFAT is only able to do so for NO. While for NO, ME-WFAT and TDDFT are in excellent agreement, the most noticeable difference between ME-WFAT and TDDFT in the case of OCS and CH$_3$Br lies in the region around $\beta=0^\circ$. For OCS, the most likely reason is the induced polarization of the core orbitals, which is not captured by ME-WFAT, whereas for CH$_3$Br, we attribute this difference to the significant contribution of the ionization channel connecting neutral ground state and a certain cation excited state around $\beta=0^\circ$. \green{The overall qualitatively good agreement between the quasi-static WFAT and RT-TDDFT simulations indicates that the ionized electron can be treated using an adiabatic description for wavelengths as short as $800$ nm.}
\fgreen{On the other hand, one may expect that WFAT is insufficient to treat non-adiabatic processes, such as inter-channel coupling during ionization and carrier-envelope phase effects.}
\fgreen{The very efficient nature of WFAT simulations (see the timings in Table \ref{tuned tddft parametes}) also makes it a fitting candidate as a tool to perform a quick scan of the ionization properties of large molecules before more elaborate calculations or actual experiments are conducted on them.}

One direction for improvement of our results would be the inclusion of the first order correction to ME-WFAT. 
\purple{This correction accounts for the distortion of orbitals due to the field, which may be needed to correct for the OCS disagreement discussed above.} Such an effort for ME-WFAT in the tail-representation has been outlined in Ref. \cite{tr_mewfat_first_order}. \purple{
An appealing application for ME-WFAT may come to the study of particle-like charge migration \cite{sideband_hhs, cm_mode, atto_soliton}} \green{where it is shown that an initial localized hole can sustain its locality throughout the periodic dynamics. Given the coupled time-space sensitivity of such a dynamic, IR ME-WFAT opens up a promising avenue in the modeling of tunnel ionization as a probe of charge migration.}

\begin{acknowledgments}
We thank Oleg Tolstikhin for the enlightening discussion about WFAT. The author is also grateful to Takeshi Sato for his suggestions regarding the degenerate Hartree-Fock solutions of CH$_3$Br$^+$. \fgreen{We also thank Peter S\'andor and Robert Jones for their permission to use their data in our work, and Adonay Sissay for his help with some simulation parameters.} \red{This work is supported by the U.S. Department of Energy, Office of Science, Office of Basic Energy Sciences, under Award No. DE-SC0012462. Portions of this research were conducted with high performance computational resources provided by Louisiana State University \cite{lsu-hpc} and the Louisiana Optical Network Infrastructure \cite{loni}}
\end{acknowledgments}

\appendix

\section{Dyson orbital for single-determinant wave functions} \label{app:dyson_orb}
When both of the neutral and cation wave functions are a single-determinant, that is,
\begin{subequations}
   \label{eq:slater_det}
   \begin{align}
       \Psi(\mathbf X_N) =& \,
       \frac{1}{\sqrt{N!}}
       \operatorname{det}\left(
       \psi_1^a \cdots \psi_{N_a}^a \, 
       \psi_1^b \cdots \psi_{N_b}^b
       \right), \\
       \Psi^+(\mathbf X_{N-1}) =& \,
       \frac{1}{\sqrt{(N-1)!}}
       \operatorname{det}\left(
       \upsilon_1^a \cdots \upsilon_{N'_a}^a \, 
       \upsilon_1^b \cdots \upsilon_{N'_b}^b \right)
   \end{align}
\end{subequations}
then the Dyson orbital may be shown to have the form
\begin{align}
    \psi_D^{\sigma} (\mathbf r) 
    =& \,
    \int d\mathbf X_{N-1} d\sigma' \,
    (\Psi^+(\mathbf X_{N-1}))^* 
    \Psi(\mathbf X_{N-1}, \mathbf{r}\sigma') \nonumber\\
    =& \, 
    \frac{(-1)^{N+\delta_{\sigma b}N_a}}{\sqrt{N}} \mathcal{R} \, \Tilde{\psi}^{\sigma} (\mathbf r), \label{eq:dyson}
\end{align}
where
\begin{align}
    \tilde{\psi}^{\sigma} (\mathbf r) =&\,
    \sum_{i=1}^{N_{\sigma}} (-1)^i \mathcal{P}(i) \psi_i^{\sigma}(\mathbf r). \label{eq:imperfect_dyson}    
\end{align}
$\sigma$ is the ionized spin channel, that is, the spin channel of the neutral from which an electron has been removed, and $\mathcal{P}(i)$ and $\mathcal{R}$ are defined in Appendix \ref{app:determinant_coefficients}. Eq. \eqref{eq:dyson} assumes that the \fgreen{following} conditions constraining the number of electrons in each spin channel of both the neutral and cation,
\begin{align*}
    N'_{\sigma} &= N_{\sigma} - 1  , \\
    N'_{p(\sigma)} &= N_{p(\sigma)},
\end{align*}
\fgreen{are satisfied. These relations are the consequence of single determinant wave functions being an eigenstate of the $z$-component of spin angular momentum.} If not, the Dyson orbital uniformly vanishes.

\section{Various overlap integrals between neutral and cation wave functions} \label{app:determinant_coefficients}

Let $S_\sigma^{\Psi^+\Psi} \left[ a',b',\ldots | a, b, \ldots \right]$ be the spin-$\sigma$ block of the overlap matrix formed between the orbitals occupied in $|\Psi^+\rangle$ removing spin-orbitals $a',b',\ldots$ and the orbitals occupied in $|\Psi\rangle$ removing spin-orbitals $a,b,\ldots$, then
\begin{subequations}
   \label{eq:det_coeff_general}
   \begin{align}
       \mathcal{P}(i) &= \operatorname{det}\left(S_{\sigma}^{\Psi^+\Psi} \left[    \varnothing \Big| \psi_i^{\sigma} \right] \right) , \label{eq:define_pcf} \\
       \mathcal{R} &= \operatorname{det}\left(S_{p(\sigma)}^{\Psi^+\Psi} \left[    \varnothing \Big| \psi_i^{\sigma} \right] \right) , \label{eq:define_rcf}\\
       \mathcal{Q}(k,i,k') &= \operatorname{det} \left( S_{\sigma}^{\Psi^+\Psi}    \left[ \upsilon_{k'}^{\sigma} \Big| \psi_i^{\sigma}, \psi_k^{\sigma}    \right] \right) 
       (1-\delta_{ik}), \label{eq:define_qcf} \\
       \mathcal{S}(k,k') &= \operatorname{det} \left(    S_{p(\sigma)}^{\Psi^+\Psi} \left[ \upsilon_{k'}^{p(\sigma)} \Big|    \psi_i^{\sigma} , \psi_k^{p(\sigma)} \right] \right)  \label{eq:define_scf}
       ,
   \end{align}
\end{subequations}
where the notation $\varnothing$ means that no orbitals are removed \fgreen{from the corresponding charge state.} 
\purple{A full derivation on how these determinants arise within the formulation of IR ME-WFAT will be presented in Ref. \cite{ir_mewfat}.}

\section{Exchange term in the ME-WFAT integral} \label{app:exchange_mewfat}
At first glance, it might be tempting to identify the $\hat K_{k'k}^\sigma$ term in Eq. \eqref{eq:v2e_term} as the exchange term.
In this section we will first show that this is not the case, and then analyze Eq. \eqref{eq:v2e_term} to identify the most suitable term to be regarded as the exchange term in the context of IR ME-WFAT using DFT Kohn Sham orbitals. As a guide, we use ME-WFAT with HF and in the case of unrelaxed cation orbitals, which are just taken from $N-1$ occupied orbitals of the neutral. We also assume the only neutral orbital unoccupied in the cation to be $\psi_{i'}^\sigma$. In this case, 
\begin{subequations}
   \label{eq:det_coeff_unrelaxed}
   \begin{align}
       \mathcal P(i) &= \delta_{ii'}, \\
       \mathcal R &= 1, \\
       \mathcal Q(k,i,k') &= (1-\delta_{ik})(\delta_{ii'} \delta_{k'\Bar k} + \delta_{ki'} \delta_{k' \Bar i}), \\
       \mathcal S(k,k') &= \delta_{k'k}.
   \end{align}
\end{subequations}
Using Eq. \eqref{eq:det_coeff_unrelaxed} in the triple sum of Eq. \eqref{eq:v2e_term}, one obtains
\begin{widetext}
   \begin{align}
       \label{eq:v2e_term_oe}
       \sum_{k'=1}^{N_{\sigma}-1} \, \sum_{j=2}^{N_{\sigma}} \, \sum_{k=1}^{j-1} 
       \bigg( 
        \hat J_{k'k}^{\sigma} |\psi_j^{\sigma} \rangle  -
        \hat K_{k'k}^{\sigma} |\psi_j^{\sigma} \rangle 
        \bigg)
        =&\,
       \sum_{k'=1}^{N_{\sigma}-1} \, \sum_{k=1}^{i'-1} (-1)^{k+k'} \delta_{k'k}
       \left(\hat V_{k'k}^{\sigma} \left| \psi_{i'}^{\sigma} \right\rangle
       - 
       \hat V_{k'i'}^{\sigma} \left| \psi_k^{\sigma} \right\rangle \right)  \nonumber \\
       &\, 
       + \sum_{k'=1}^{N_{\sigma}-1} \, \sum_{k=i'+1}^{N_{\sigma}} \,  (-1)^{k+k'}    \delta_{k', k-1}
       \left( \hat V_{k'i'}^{\sigma} \left| \psi_k^{\sigma} \right\rangle 
       - 
       \hat V_{k'k}^{\sigma} \left| \psi_{i'}^{\sigma} \right\rangle
       \right)  \nonumber \\
       =&\,
       \sum_{k\neq i'}^{N_{\sigma}}
       \left(\hat V_{kk}^{\sigma} \left| \psi_{i'}^{\sigma} \right\rangle
       - 
       \hat V_{ki'}^{\sigma} \left| \psi_k^{\sigma} \right\rangle \right),
   \end{align}
\end{widetext}
where we note that the last expression is just the classical repulsion and exchange terms among electrons with the same spin in the OE-WFAT integral formula. 

A closer look at the first and second lines of the right-hand side reveals that there are certain pairs of $(j,k)$ from both the $\hat J_{k'k}^\sigma$ and $\hat K_{k'k}^\sigma$ terms on the left-hand side that lead to the exchange (second) term in the third line. In particular, on the right-hand side, the second term inside the parentheses of the first line plus the first term inside the parentheses of the second line together constitute the exchange term in the third line. This is why in general one cannot assign $\hat K_{k'k}^\sigma$ alone as the exchange in the ME-WFAT integral. We can, however, identify an occupied orbital of the neutral that is 'most similar' to the Dyson orbital. The most straightforward method of this identification is by finding the largest absolute value of the elements of $\mathcal P(i)$, which, in most cases, corresponds to $\psi_{i'}^\sigma$. Then by reordering the neutral orbitals such that this largest element is at $N_\sigma$ (that is, $i'=N_\sigma$), we can eliminate the second line of Eq. \eqref{eq:v2e_term_oe}. After this reordering, one can identify the $\hat K_{k'k}^\sigma$ term in the left-hand side of this equation as the exchange contribution.

\providecommand{\noopsort}[1]{}\providecommand{\singleletter}[1]{#1}%
%


\end{document}